\documentclass[final,3p,times,authoryear]{elsarticle}

\usepackage{hyperref}
\usepackage{amssymb}
\usepackage{amsthm}
\usepackage{amsmath}
\usepackage{scalerel}
\usepackage{dcolumn}
    \newcolumntype{d}[1]{D{.}{.}{#1}}

\usepackage{pgfplots}
\usepackage{tikz}
\usetikzlibrary{pgfplots.external,}
\pgfplotsset{compat=1.12}
\tikzset{every picture/.style=semithick}
\usepackage{pgfgantt}
\usepackage{pdflscape}%
\usetikzlibrary{plotmarks}
\newlength\fheight 
\newlength\fwidth 
\DeclareRobustCommand\sampleline[1]{%
  \tikz\draw[#1] (0,0) (0,\the\dimexpr\fontdimen22\textfont2\relax)
  -- (2em,\the\dimexpr\fontdimen22\textfont2\relax);%
}

\usepackage{import}
\usepackage{booktabs}
\RequirePackage{xcolor}
\usepackage{subcaption}
\graphicspath{{figures/svg_export/}}

\newcommand{\partialdiff}[2]{ \frac{\partial {#1}}{\partial {#2}} }

\global\long\def\tns#1{{\boldsymbol{#1}}}
\global\long\def\vct#1{\boldsymbol{#1}}

\journal{Computers \& Fluids}
\begin{document}

\begin{frontmatter}

\title{Adaptive Reduced-Order Modeling for Non-Linear Fluid-Structure Interaction}

\author[x1]{Ali Thari}
\author[x2]{Vito Pasquariello}
\author[x3]{Niels Aage}
\author[x1]{Stefan Hickel \corref{cor1}}
\ead{S.Hickel@tudelft.nl}
\cortext[cor1]{Corresponding author.}

\address[x1]{Aerodynamics Group, Faculty of Aerospace Engineering, Technische Universiteit Delft, Kluyverweg~1, 2629 HS Delft, The Netherlands}
\address[x2]{Lilium GmbH, Claude-Dornier Str. 1, 82234 Wessling, Germany}
\address[x3]{Department of Mechanical Engineering, Technical University of Denmark, DK-2800 Lyngby, Denmark}

%!TEX root = ../AROM_v3.tex

\begin{abstract}
We present an adaptive reduced-order model for the efficient time-resolved simulation of fluid-structure interaction problems with complex and non-linear deformations. The model is based on repeated linearizations of the structural balance equations. Upon each linearization step, the number of unknowns is strongly decreased by using modal reduction, which leads to a substantial gain in computational efficiency.  Through adaptive re-calibration and truncation augmentation whenever a non-dimensional deformation threshold is exceeded, we ensure that the reduced modal basis maintains arbitrary accuracy for small and large deformations. Our novel model is embedded into a partitioned, loosely coupled finite volume - finite element framework, in which the structural interface motion within the Eulerian fluid solver is accounted for by a conservative cut-element immersed-boundary method. Applications to the aeroelastic instability of a flat plate at supersonic speeds, to an elastic panel placed within a shock tube, and to the shock induced buckling of an inflated thin semi-sphere demonstrate the efficiency and accuracy of the method.
\end{abstract}


\end{frontmatter}

% include line numbers for reviewing purpose
%\linenumbers
\setlength{\baselineskip}{20pt}
 
%!TEX root = ../AROM_v3.tex

\section{Introduction}\label{sec:intro}

Fluid-Structure Interaction (FSI) occurs in a broad range of applications, such as blood flow though heart valves \citep{Peskin1}, flutter of aircraft wings \citep{Farhat1998} and shock-induced deformations of rocket nozzles and panels \citep{Garelli, vito_conf}. FSI simulations involve two different branches of computational physics: Computational Fluid Dynamics (CFD), which is often based on an Eulerian finite-volume representation, and Computational Solid Mechanics (CSM), for which a finite-element discretization is frequently chosen.

FSI algorithms can be divided into monolithic and partitioned methods. The monolithic approach is characterized by solving a single set of discrete equations describing entire coupled system \citep{Kloppel}. While this procedure may be time-consuming, it is robust, accurate and stable.
On the other hand, a partitioned approach is frequently used to conveniently couple off-the-shelf CFD and CSM codes, which are extensively validated and employ the most efficient numerical schemes for each discipline. Partitioned methods can be further classified as strongly or loosely coupled, where the distinction is based upon whether or not the complete set of coupling conditions at the conjoined FSI interface is satisfied. For similar mass density of fluid and solid, loosely coupled methods may suffer from the artificial added-mass effect, possibly leading to computational instabilities \citep{Farhat1998,Causin}. Stability can be recovered by introducing sub-iterations \citep{Kuttler}, which however increase the computational cost significantly \citep{Farhat2004}. \citet{Badia} obtained very promising results by employing a Robin-type boundary condition at the FSI interface. Similarly, \citet{Banks2,Banks3} introduced so-called Added-Mass Partitioned algorithms to overcome the added-mass effect for incompressible flow as well as for very light rigid bodies in compressible flow. For the majority of FSI applications, such as compressible aeroelasticity, however, a loosely coupled method is sufficient \citep{Farhat1998}.

In practice, one of the main challenges is to keep the computational cost of time-resolved FSI simulations at a reasonable level without sacrificing the required accuracy. 
The present paper addresses the main performance bottleneck of high-fidelity turbulence resolving FSI simulations with two loosely coupled domain-specific codes, where the time step size is restricted by the resolution requirements of the fluid flow: CFD solvers for Large Eddy Simulations (LES) and Direct Numerical Simulations (DNS) usually employ high-order finite-volume or finite-difference space discretizations and explicit time marching methods, which efficiently satisfy the high resolution requirements of the fluid flow, and can be parallelized easily with excellent scalability on massively-parallel supercomputers. The parallelization of finite-element CSM methods is less straight-forward; in our experience, most parallelized CSM solvers require a shared memory architecture and usually show limited parallel scalability. This leads to the curious situation that one time step for advancing the orders of magnitude more expensive (in terms of CPU time) CFD equations requires much less run time (wall-clock time) than one time step of the CSM problem \citep{vito_conf,Vito}. 

This does not mean that finite-element CSM methods are per se inefficient; they are designed for much larger time steps, which would be effective in a stand-alone setting. \citet{PIPERNO} addressed this bottleneck with a sub-cycling approach, where the fluid solver is advanced multiple times before the structural solution is advanced in one large time step. The efficiency and stability of various sub-cycling methods are discussed by \citet{Farhat_sub}. As an alternative, Reduced-Order Modeling (ROM) can be used to improve the efficiency of the CSM solver \citep{Dowell1996,Hall,Rixen}. One of the first reported model reductions was presented by \citet{Rayleigh}, who employed the Mode Superposition Method (MSM) to approximate the displacement field with a low number of free vibration modes. The method truncates the vibration modes of the structure at low number, i.e.~$N<<n$, where $n$ is the number of degrees of freedom and $N$ is the number of dynamically important modes. Several improvements for truncated modal superpositions have been proposed since then, one being the Static Correction Method (SCM) used by \citet{Rixen}, \citet{Wilson} and \citet{Besselink}, among others. This method accounts for the omitted modes by including the truncated modes statically, which leads to a more adequate representation of the modal loads. However, this method is only effective if the structure has very low natural frequencies. \citet{Dickens} introduced the Modal Truncation Augmentation (MTA) method for the dynamic correction of the load representation.
MTA improves the overall accuracy compared to the SCM and is also effective for a broader frequency range \citep{Dickens}.
Another popular reduction method is the Craig-Bampton Method (CBM) presented in \cite{Bampton} and the family of Component Mode Synthesis (CMS), see, e.g., \citet{Qu}. These methods divide the global finite-element structure into several sub-structures connected with an appropriate interface description. The CMS method is also known as Super Element Method in the sense that each substructure can be considered as a single finite element.

Linear ROM approaches generally fail for non-linear problems. Some of the earliest reported work on non-linear ROM was presented by \citet{Morris}.
It follows a modal superposition that incorporates the stiffness matrix at a deformed state and was applied for frame structures. Similarly, \citet{Remseth} employed a Taylor linearization of the reduced-order finite-element equations with respect to a deformed state to account for non-linear effects on frame structures. 
The projection on a new eigenmode basis (around a new deformed state) generates truncation errors, which can accumulate.
\citet{Nickell} used a Rayleigh-Ritz analysis to derive a reduction model that includes non-linear effects through modal derivates along with the eigenmodes and exploits
those derivates along with tangent modes to limit the need of eigenspace updates. Alternative methods that minimize the need to update the reduction basis have been presented by \citet{Idelsohn} \& \citet{Tiso}. \citet{Mignolet} reviewed ROMs for non-linear geometric structures based on indirect methodologies, where the non-linear stiffness terms are approximated by cubic polynomials.
A key aspect of the ROM effort is to properly select the basis functions. The authors present a strategy that enriches the basis of free vibration modes by dual modes for capturing non-linear structural behavior. A set of non-linear static simulations with representative loads is needed to determine the dual modes, which are calculated based on the proper orthogonal decomposition of the series of non-linear displacement fields.
Recently, \cite{Wu} presented a ROM for flexible multi-body systems with large non-linear deflections, where the ROM is based on a combination of the CBM and cubic polynomials of the configuration dependent terms.

In this paper we derive an adaptive ROM (AROM) based on adaptive re-calibration of the reduced modal basis with MTA correction, which allows us to maintain arbitrary accuracy also in the case of large and non-linear structural deformations.
The AROM is imbedded into the loosely coupled partitioned FSI algorithm proposed by \citet{Vito}.
We employ an established high-fidelity turbulence resolving finite-volume method for solving the three-dimensional compressible Navier-Stokes equations on block-structured adaptive Cartesian grids (INCA, \url{https://inca-cfd.org}) and an unstructured finite-element method for the discretization of the structural domain (CalculiX, \url{http://www.calculix.de}).
The time-varying fluid-structure interface is accounted for by the cut-element based Immersed Boundary Method (IBM) that was introduced by \citet{Orley} and then extended to deformable structures and compressible FSI applications by \citet{Vito}. 
The CFD solver is designed to provide (close to) ideal parallel scalability on thousands of cores \citep{inca_par}. Due the small time step size imposed by the resolution requirements of compressible fluid flows and the necessarily synchronous nature of alternating between the CFD and CSM at each time step, the CSM solver constitutes the critical run-time bottleneck in this framework. The essential original contribution of this work is the development and demonstration of a novel FSI-AROM algorithm, which is capable of handling structures with large, non-linear deformations accurately with high computational efficiency. 

The paper is structured as follows: The governing equations of the fluid and structure and the numerical formulation are presented in Sections \ref{sec:mathmodel} and \ref{sec:nummodel}. The coupling algorithm for non-matching time-varying interfaces is presented in Section \ref{sec:coupling}. The novel FSI-AROM method is derived in Section \ref{sec:AROM}. In Section \ref{sec:2d_prob} we validate the FSI-AROM method for linear and non-linear problems, and demonstrate the prediction capabilities for flow-induced buckling of a three-dimensional thin semi-spherical shell. A final discussion and concluding remarks are given in Section \ref{sec:conclusion}.

%!TEX root = ../AROM_v3.tex

\section{Governing Equations}\label{sec:mathmodel}

The domain of interest $\Omega = \Omega_F \cup \Omega_S $ is divided into non-overlapping fluid $\Omega_F$ and solid $\Omega_S$ subdomains with a conjoined interface \mbox{$\Gamma = \Omega_F \cap \Omega_S$}. The interface normal vector $\vct{n}^\Gamma$ is assumed to point into the fluid domain. In the following a brief description of the mathematical models required for both subdomains is given. Unless specified otherwise, we use the Einstein summation convention.

\subsection{Fluid}

The fluid flow within the domain $\Omega_F$ is governed by the three-dimensional compressible Navier-Stokes equations 
\begin{equation}\label{eq:ns}
	\frac{\partial \vct{w}}{\partial t} +\nabla \cdot \tns{H\left( w \right)}  = \vct{0}\,,
\end{equation}
which describe the conservation of mass, linear momentum and total energy. We use Cartesian coordinates where $\nabla = \left(\frac{\partial }{\partial x_1}, \frac{\partial }{\partial x_2},\frac{\partial }{\partial x_3}\right)$. The state vector $\vct{w}$ and flux tensor $\tns{H\left( w \right)} = \left[\vct{H}^{(1)}, \vct{H}^{(2)}, \vct{H}^{(3)} \right] $ are given as
\begin{equation}
	\vct{w} =
	\left[
	\begin{array}{c}
	    	\rho_F 
		\\ \rho_F u_1
		 \\ \rho_F u_2 
		 \\ \rho_F u_3 \\ \rho_F e_t
	\end{array}
	\right],  \quad
	\vct{H^{(i)}\left( w \right)} = 
	\left[
	\begin{array}{c}
			u_i \rho_F 
			\\ u_i \rho_F u_1 + \delta_{i1} p - \tau_{i1}
			\\ u_i \rho_F u_2 + \delta_{i2} p - \tau_{i2}
			\\ u_i \rho_F u_3 + \delta_{i3} p - \tau_{i3}
			\\ u_i \left(\rho_F e_t + p\right) - u_k \tau_{ik} + q_i
	\end{array}
	\right]\, ,
\end{equation}
where $u_i$ is the velocity, $\rho_F$ the fluid density, and $\rho_Fe_t$ is the total energy density.
The viscous stress tensor for a Newtonian fluid is  
\begin{equation}
	\tau_{ij} = \mu_F \left(\frac{\partial u_i}{\partial x_j} + 
	\frac{\partial u_j}{\partial x_i} \right) + \lambda_F  \frac{\partial u_k}{\partial x_k} \delta_{ij} \, ,
\end{equation}
where the first Lam\'e parameter is related to the dynamic viscosity $\mu_F$ according to Stoke's hypothesis: $\lambda_F = - {2}/{3} \mu_F$.
The heat flux is evaluated according to Fourier's law,
\begin{equation} 
	q_i = -k \partialdiff{T}{x_i}\, ,
\label{eq:four}
\end{equation}
with the  coefficient of thermal conductivity $k$.
We consider air as a perfect gas with $\gamma = 1.4$ and specific gas constant of $R=287.058 \,\frac{J}{kg\cdot K}$.
The pressure $p$ and temperature $T$ are calculated from the definition of total energy 
\begin{equation}\label{eq:energy}
	\rho_Fe_t = \frac{1}{{\gamma} - 1}p + \frac{1}{2	} \rho_F u_i u_i \, 
\end{equation}
and the ideal-gas equation of state
\begin{equation}
	p = \rho_F R T\, .
	\label{eq:idealgas}
\end{equation} 

\subsection{Solid}

The governing equations for the solid are based on the local form of the balance of linear momentum
\begin{equation}
	\rho_{S;0} \frac{\partial^2 \vct{d}}{\partial t^2} = \nabla_0\cdot \tns{P}  + \hat{\vct{b}}_0 \, , %\in \Omega_S\, ,
	\label{eq:struc4}
\end{equation}
which describes an equilibrium between the work done by inertia, internal and external forces expressed in the underformed configuration.
The vector of displacements is denoted by $\vct{d}$, $\rho_{S;0}$ is the material density of the solid, $\nabla_0 \cdot \left( \: \right)$ is the material divergence operator, $\tns{P} = \tns{F} \cdot \tns{S}$ is the first Piola-Kirchhoff stress tensor, where $\tns{F}$ is the deformation gradient tensor, 
and external material body forces are represented by $\hat{\vct{b}}_0$. 
The second Piola-Kirchhoff stress tensor is
\begin{equation}
	\tns{S} = \frac{\partial \Psi}{\partial \tns{E}}\, .
	\label{eq:second_piola_stress}
\end{equation}
In this work, a hyperelastic Saint Venant-Kirchhoff material model is chosen. 
Its associated strain energy density function $\Psi$ is given as
\begin{equation}
	\Psi\left(\tns{E}\right) = \mu_S \tns{E}:\tns{E} + \frac{1}{2} \lambda_S\left(\tns{E}:\tns{I}\right)^2\, ,
\end{equation}
where $\lambda_S$ and $\mu_S$ are the first and second Lam\'e parameter. The Green-Lagrange strain tensor is defined as
\begin{equation}
	\tns{E} = \frac{1}{2}\left(\tns{F}^T\cdot \tns{F} - \tns{I}\right) = \frac{1}{2}\left(\tns{D} + \tns{D}^T +\overbrace{ \tns{D}^T\cdot \tns{D}}^{\text{non-linear} }\right)\, ,
	\label{eq:Green-Lag}
\end{equation}
and $\tns{D} $ is the displacement gradient tensor.
The Cauchy stress tensor $\tns{\sigma}_S$, also called true stress tensor, is defined as 
\begin{equation}
	\tns{\sigma}_S = \frac{1}{J} \tns{P} \cdot \tns{F}^T\, ,
	\label{eq:cauchy_stress}
\end{equation}
where $J$ denotes the Jacobian determinant.

Boundary conditions need to be specified on $\partial \Omega_S = \Gamma_{S,D} \cup \Gamma_{S,N} \cup \Gamma$ to make the system \eqref{eq:struc4} well posed. Two different types are considered in this work, namely Dirichlet $\Gamma_{S,D}$ and Neumann $\Gamma_{S,N}$ boundaries for which we either prescribe displacements $\vct{\hat{d}}$ or tractions $\vct{\hat{t}}$
\begin{equation}
	\vct{d} = \vct{\hat{d}}  \; \text{on~} \Gamma_{S,D}\; \text{and} \; 
	\tns{P} \cdot \vct{n}_0 = \vct{\hat{t}}  \; \text{on~} \Gamma_{S,N}\, .
\end{equation}
Here $\vct{n}_0$ denotes the unit normal vector in material configuration. Further, initial conditions for displacements and velocities must be specified:
\begin{equation}
	\vct{d}^0 = \vct{d}\left(t=0\right)=\vct{\hat{d}}^0
	\; \text{on~} \Omega_{S} \;  \text{and} \; 
	\vct{\dot{d}}^0 = \vct{\dot{d}} \left(t=0\right)=\vct{\hat{\dot{d}}} ^0 \; \text{on~} \Omega_{S} \, .
\end{equation}

\subsection{Fluid-structure interface conditions}

The interface between fluid and structure requires coupling conditions. 
Tractions on $\Gamma$ have to be in equilibrium, that is,
\begin{equation}
	\tns{\sigma}_S^{\Gamma} \cdot  \vct{n}^{\Gamma}=
	\tns{\sigma}_F^{\Gamma} \cdot  \vct{n}^{\Gamma} \, .
	\label{eq:intf_cond_tractions}
\end{equation} 
Herein, $\tns{\sigma}_S$ is the Cauchy stress tensor given by Eq.~\eqref{eq:cauchy_stress} and 
\begin{equation}
	\tns{\sigma}_F = -p \tns{I} + \tns{\tau}
\end{equation} 
denotes the fluid stress tensor comprising an inviscid and viscous contribution. In addition, the  kinematic no-slip boundary condition
\begin{equation}
	\frac{\partial \vct{d}^{\Gamma}}{\partial t} = \vct{u}^{\Gamma}\, 
	\label{eq:intf_cond_kinematic_viscous}
\end{equation}
 must be satisfied, which in case of an inviscid flow reduces to matching normal velocities on $\Gamma$
\begin{equation}
	\frac{\partial \vct{d}^{\Gamma}}{\partial t} \cdot \vct{n}^{\Gamma} = \vct{u}^{\Gamma}\cdot  \vct{n}^{\Gamma}\, .
	\label{eq:intf_cond_kinematic_inviscid}
\end{equation}
%!TEX root = ../AROM_v3.tex

\section{Numerical models}\label{sec:nummodel}

\begin{figure}[t]
  \centering
  \includegraphics[]{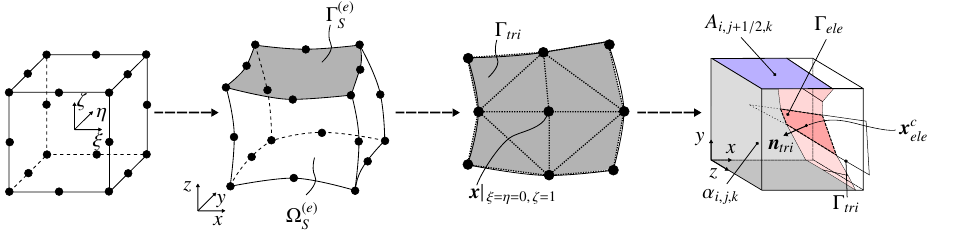}
  \caption{Schematic triangulation of a structural interface element $\Gamma_S^{(e)}$. Resulting interface triangles $\Gamma_{tri}$ are used as an input for the cut-algorithm to compute individual cut-elements $\Gamma_{ele}$ and cut-cell related geometric quantities.}
  \label{fig:triangulation_interface}
\end{figure}

\subsection{Fluid}\label{sec:num_fluid}

The compressible Navier-Stokes equations, Eq.~\eqref{eq:ns}, are discretized with a finite volume method based on the integral form
\begin{equation}
	\int_{t^n}^{t^{n+1}} \int_{\Omega_{i,j,k}\cap \Omega_F} \frac{\partial \vct{w}}{\partial t} dV\, dt +  
	\int_{t^n}^{t^{n+1}} \int_{\partial\left(\Omega_{i,j,k}\cap \Omega_F\right)} \tns{H}(\vct{w}) \cdot \vct{n} \, dS\,\,dt = 0\, ,
	\label{eq:NS2}
\end{equation}
where Gauss's theorem has been applied. The integral is taken over $\Omega_{i,j,k}\cap \Omega_F$, i.e., the part of a Cartesian computational cell $\Omega_{i,j,k}$ that belongs to the fluid domain $\Omega_F$, and over the time step $\Delta t = t_{n+1} - t_{n}$. 
In order to account for the FSI interface within the fluid solver, which operates on Cartesian grids, we employ the cut-element IBM of \citet{Orley} and \citet{Vito}. 
The discrete FSI interface is composed of several structural interface elements $\Gamma_S^{(e)}$. Each structural interface element $\Gamma_S^{(e)}$ is triangulated as shown exemplarily in Fig.~\ref{fig:triangulation_interface} for a quadratic hexahedral element. The resulting interface triangles $\Gamma_{tri}$ are used as an input for the IBM algorithm. A fluid cell that is cut by at least one interface triangle $\Gamma_{tri}$ is referred to as a cut-cell. The fluid-solid interface within a cut-cell is composed of one or several cut-elements $\Gamma_{ele} = \Gamma_{tri} \cap \Omega_{i,j,k} $, each representing one or a part of one interface triangle, see Fig.~\ref{fig:triangulation_interface}. 
Applying a volume average of the conserved variables
\begin{equation}
	\overline{\vct{w}}_{i,j,k}=\frac{1}{\alpha_{i,j,k}V_{i,j,k}}\int_{\Omega_{i,j,k}\cap \Omega_F} \vct{w}\, dx \, dy \, dz\, ,
	\label{NS3}
\end{equation} 
and considering (for demonstration purposes) a simple forward Euler time integration scheme leads to the following discrete form of Eq.~\eqref{eq:NS2}

\begin{align}
	\alpha_{i,j,k}^{n+1} \overline{\vct{w}}_{i,j,k}^{n+1} = \alpha_{i,j,k}^{n} \overline{\vct{w}}_{i,j,k}^{n} 
	&+ \frac{\Delta t}{\Delta x_i}\Big[A_{i-1/2,j,k}^n\tns{H}^{(1)}_{i-1/2,j,k} - 
	A_{i+1/2,j,k}^n\tns{H}^{(1)}_{i+1/2,j,k}\Big] \nonumber \\ 
	&+ \frac{\Delta t}{\Delta y_j}\Big[A_{i,j-1/2,k}^n\tns{H}^{(2)}_{i,j-1/2,k} - 
	A_{i,j+1/2,k}^n\tns{H}^{(2)}_{i,j+1/2,k}\Big] \nonumber \\
	&+ \frac{\Delta t}{\Delta z_k}\Big[A_{i,j,k-1/2}^n\tns{H}^{(3)}_{i,j,k-1/2} - 
	A_{i,j,k+1/2}^n\tns{H}^{(3)}_{i,j,k+1/2}\Big] \nonumber \\
	&+ \frac{\Delta t}{V_{i,j,k}}\vct{\chi}_{i,j,k}\ .
	\label{NS4}
\end{align}
Herein $\alpha_{i,j,k}$ is the fluid volume fraction of the cut-cell, $V_{i,j,k} = \Delta x_i \Delta y_j \Delta z_k $ is the total volume of cell $\Omega_{i,j,k}$ and $A$ is the effective fluid wetted cell-face aperture, see also Fig.~\ref{fig:triangulation_interface}. The face-averaged numerical fluxes across the regular cell faces are $\tns{H}^{(i)}$ and $\vct{\chi}_{i,j,k} = \sum_{ele} \vct{\chi}_{ele}$ denotes the integral flux across the interface $\Gamma_{i,j,k} = \sum_{ele} \Gamma_{ele}$, where the latter is only present for a cut-cell. 
The interface fluxes $\vct{\chi}_{ele}$ include the fluid stresses due to pressure and viscous effects, the resulting work at the interface, and heat transfer through the interface. For a detailed description of the cut-element IBM please refer to \cite{Orley} and \cite{Vito}.

For the spatial reconstruction and numerical flux functions we either use the Adaptive Local Deconvolution Method (ALDM) by \citet{Hickel1,Hickel2}, or the $5^{th}$-order WENO (Weighted Essentially Non-Oscillatory) scheme by \citet{Liu} with the HLLC flux \citep{Toro1994}. In order to avoid modified interpolation stencils in the FV reconstruction near the interface, we assign special ghost fluid states that depend on the interface boundary conditions to non-cut fluid cells within the solid part of the domain \citep{Mittal1,Vito}. Finally, time integration is performed with a conditionally stable, explicit third-order Runge-Kutta scheme.

\subsection{Solid}\label{sec:num_solid}

We cast the structural equations, Eq.~\eqref{eq:struc4}, into their weak form by applying the principle of virtual work with virtual displacements $\delta \vct{d}$ and subsequently integrating the balance equation over the structural subdomain $\Omega_S$. Following this procedure and applying Gauss's theorem yields
\begin{equation}
	\int_{\Omega_{S}}\left( \rho_{S;0} \ddot{\vct{d}} \cdot \delta \vct{d} + \tns{S}:\delta\tns{E} -  \vct{\hat{b}}_0 \cdot \delta \vct{d} \right) dV_0 
	- \int_{\Gamma_{S,N}} \hat{\vct{t}}_0 \cdot \delta \vct{d}\,dA_0 - \delta W_{S}^{\Gamma} = {0}\, ,
	\label{eq:weakform}
\end{equation}
where $dA_0$ and $dV_0$ are infinitesimal surface and volume elements, respectively, and $\delta \tns{E}$ is a result of the variation of the strain expression in Eq.~\eqref{eq:Green-Lag},
\begin{equation}
	\delta \tns{E} = \frac{1}{2	} \left(\tns{F}^{T}\delta \tns{D} + \delta \tns{D}^T \tns{F}\right)\, .
	\label{eq:virt_strain}
\end{equation}
The weak form, Eq.~\eqref{eq:weakform}, represents the balance of virtual work $\delta W$, namely
\begin{equation}
	\delta W_{inertia} + \delta W_{internal} - \delta W_{body forces} - \delta W_{traction} - \delta W_S^{\Gamma}  = 0\, ,
\end{equation}
where the work at the FSI interface is $\delta W_S^{\Gamma}$. 
We use the FEM to discretize the integral equation \eqref{eq:weakform} in space. The structural domain $\Omega_S$ is composed of $n^e$ solid elements $\Omega_S^{(e)}$ with consistent basis functions for representing the displacement field. 
The semi-discrete form of Eq.~\eqref{eq:weakform} is then obtained by assembling the contributions of all elements $\Omega_S^{(e)}$, resulting in
\begin{equation}
\tns{M}\ddot{\vct{d}} + \vct{f}_{S;int}(\vct{d}) - \vct{f}_{S;ext}(\vct{d})- \vct{f}_{S}^{\, \Gamma} = \vct{0}\, ,
\label{eq:semi_str_eq}
\end{equation}
with the mass matrix $\tns{M}$, the discrete acceleration vector $\ddot{\vct{d}}$ and the discrete displacement vector $\vct{d}$. The forces are divided into internal forces $ \vct{f}_{S;int}$, external forces $\vct{f}_{S;ext}$ and interface forces $\vct{f}_S^{\,\Gamma}$ resulting from the fluid. 

In contrast to \citet{Vito}, who used linear FE together with element technology based on Enhanced Assumed Strains to avoid locking phenomena, we use quadratic shape functions for interpolating the displacements on $\Omega_S^{(e)}$ unless stated otherwise. 
The final step is to discretize Eq.~\eqref{eq:semi_str_eq} in time. We employ the Hilber-Hughes-Taylor $\alpha$-method \citep{Hilber1977} for time integration. Due to its implicit character a coupled set of non-linear equations needs to be solved, which is done by the Newton-Raphson method. 

%!TEX root = ../AROM_v3.tex

\section{Coupling methods}\label{sec:coupling}

\subsection{Load and motion transfer}\label{sec:load_transfer}

The cut-cell discretization inevitably leads to non-matching grids at the conjoined interface $\Gamma$ and requires interpolation methods for the load transfer between both subdomains. Specifically, we search for the discrete force vector $\vct{f}_S^\Gamma$ that results from the fluid tractions acting on the wetted structure. We follow the approach suggested by \cite{Farhat1998_LOAD} and use the shape functions of $\Omega_S^{(e)}$ for interpolating the fluid loads on the adjacent structural nodes. Consider a single cut-element $\Gamma_{ele}$ as shown in Fig.~\ref{fig:triangulation_interface}. The fluid forces  $\vct{f}_F^{ele}$ follow directly from the pressure and viscous contributions to the IBM interface flux $\vct{\chi} = \sum_{ele} \vct{\chi}_{ele}$ in Eq.~\eqref{NS4}. Since the structural interface, i.e., the triangulation $\Gamma_{tri}$, directly serves as an input for the IBM, there is no extra need for a pairing algorithm to associate a single face centroid $\vct{x}_{ele}^c$ to the closest wet structural interface segment $\Gamma_S^{(e)}$. However, we have to determine the natural coordinates $\vct{\xi}_{ele}^c(\vct{x}_{ele}^c)$ of this fluid point. This inverse mapping problem is solved iteratively with a Newton-Raphson method. Finally, the force contribution from a single cut-element to an individual node of the paired structural interface segment $\Gamma_S^{(e)}$ is given by
\begin{equation}
 \vct{f}_{S,k}^{\Gamma} = N_k(\vct{\xi}_{ele}^c) \, \vct{f}_F^{ele}\, ,
 \label{eq:interpol_loads}
\end{equation}
where $N_k$ denotes the shape function of the $k-$th structural node on $\Gamma_S^{(e)}$. Summing up the contributions of all cut-elements in $\Omega_F$ leads to the interface force vector $\vct{f}_S^\Gamma$. It is easy to verify that this interpolation guarantees a global conservation of loads over the interface by recalling that all shape functions at one specific location sum up to unity.

The cut-element IBM requires the velocity at the face centroid $\vct{x}_{ele}^c$ for evaluating the work done at the interface. We use the same interpolation strategy based on the shape functions of the structural domain
\begin{equation}
 \vct{u}^{\Gamma;ele} = \sum_{k \, \in \, \Gamma_S^{(e)}} N_k(\vct{\xi}_{ele}^c) \, \dot{\vct{d}}_k \, ,
 \label{eq:interpol_velo}
\end{equation}
where $\dot{\vct{d}}_k$ is the velocity of the $k-$th structural node on $\Gamma_S^{(e)}$.

The motion of the structure within the fluid domain is accounted for by updating the cut-elements (and cut-cells) after each time step based on the triangulated interface $\Gamma_{tri}$ \citep{Vito}. Consequently the compatibility between the displacement fields of the structure and the fluid at the FSI interface is implicitly fulfilled in a discrete sense for all structural nodes $k \in \Gamma_S$ and no further interpolation is required.

\subsection{Summary of the coupling procedure}\label{sec:sum_coup}

We use an explicit, first-order in time accurate, loosely coupled FSI algorithm to advance the system from time level $t^n$ to $t^{n+1} = t^n + \Delta t^n$. 
The main steps are summarized below:
\begin{enumerate}
\item At time level $t^n$ the structural displacements $\vct{d}^{\Gamma;n}$ and velocities $\dot{\vct{d}}^{
\Gamma;n}$ at the interface are used to update the cut-cell geometry. The triangulated interface $\Gamma_{tri}$, see Fig.~\ref{fig:triangulation_interface}, is used as an input for the cut-algorithm. 
\item The fluid equations, Eq.~\eqref{NS4},  are advanced in time. The interface exchange term, 
as well as the ghost-cell methodology use known structural quantities at time level $t^n$. The interpolation of the structural interface velocities to the cut-elements is described in Section \ref{sec:load_transfer}.
\item The newly computed fluid interface tractions $\tns{\sigma}_F^{\Gamma; n+1} \cdot \vct{n}^{\Gamma;n}$ are projected to the structural interface elements as described in Section \ref{sec:load_transfer} with the help of the shape functions of $\Omega_S^{(e)}$.
\item The structural system,  Eq.~\eqref{eq:semi_str_eq}, is solved and advanced in time with the projected fluid tractions from time level $t^{n+1}$ imposed as additional Neumann boundary conditions. 
\item Proceed to the next time step.
\end{enumerate}
%!TEX root = ../AROM_v3.tex

\section{Adaptive Reduced-Order Model}\label{sec:AROM}

\subsection{Linearization and modal truncation}

In this section, we propose a numerical framework for switching between a full FEM description and a more efficient Adaptive Reduced-Order Model (AROM) that maintains accuracy also when a structure undergoes large non-linear deformations. The algorithm is based on Taylor expansion around a reference state $\vct{d}_{ref}$. Linearizing Eq.~\eqref{eq:semi_str_eq} around this reference leads to
\begin{equation}
	\tns{M}\ddot{\vct{d}}_{\text{ref}} + \vct{f}_{S;int}(\vct{d}_{\text{ref}}) - \vct{f}_{S;ext} - \vct{f}_S^{\, \Gamma} +  
	\tns{M} \left( \ddot{\vct{d}} - \ddot{\vct{d}}_{\text{ref}} \right)
 	+ \frac{\partial \vct{f}_{S;int}(\vct{d}_{\text{ref}})}{\partial \vct{d}}  \left( \vct{d} - \vct{d}_{\text{ref}} \right) 
 	= \vct{0}\, .
 	\label{eq:linearized}
\end{equation}
Notice that the reference state can be either a given initial condition or the FEM solution at the switching point between classical FEM and AROM. 
We introduce a new variable, $\delta \vct{d} =  \vct{d} -  \vct{d}_{\text{ref}}$ ,
for the deflection with respect to the reference state $\vct{d}_{\text{ref}}$. Consequently, time derivatives of $\delta \vct{d}$ reduce to 
\begin{equation}
	\delta \dot{\vct{d}} =  \dot{\vct{d}}  \text{~~and~~}
	\delta \ddot{\vct{d}} =  \ddot{\vct{d}}\, .
\end{equation}
Rearranging Eq.~\eqref{eq:linearized} leads to
\begin{equation}
	 \tns{M}\delta\ddot{\vct{d}}
	+  \tns{K}\left(\vct{d}_{\text{ref}}\right) \delta \vct{d} 
 	= \vct{f}_{S;ext} + \vct{f}_S^{\, \Gamma} -
 	\vct{f}_{S;int}(\vct{d}_{\text{ref}})\, ,
	 \label{eq:linear:2}
\end{equation}
where the tangent stiffness matrix $\tns{K}\left(\vct{d}_{\text{ref}}\right)$ represents the Jacobian of the internal forces
\begin{equation}
	\tns{K}\left(\vct{d}_{\text{ref}}\right)
	 = \frac{\partial \vct{f}_{S;int}(\vct{d}_{\text{ref}})}{\partial \vct{d}}\, .
 	\label{eq:upd_K}
\end{equation}
The initial conditions for $\delta \vct{d}$ are

\begin{align}
	\delta \vct{d}^0  &=\vct{d}^{{n}} - \vct{d}_{\text{ref}} = \vct{0}\, , \\
	\delta\dot{\vct{d}}^0  &= \dot{\vct{d}}^{{n}}\, , \\
	\delta\ddot{\vct{d}}^0  &= \ddot{\vct{d}}^{{n}}\, ,
\end{align}
where the superscript ${n}$ denotes the (last) results obtained with the full FEM model, Eq.~\eqref{eq:semi_str_eq}, before switching to AROM. Since this initial condition is also considered as the reference state, i.e., $\vct{d}^n = \vct{d}_{\text{ref}}$, the initial condition for the deflections $\delta \vct{d}^0$ is zero.

Equation \eqref{eq:linear:2} is expressed in the physical space. For reduced-order modeling we shrink the system of equations by the mode superposition method \citep{Rayleigh,Dowell2001}. In a first step, the eigenmodes of the structure are obtained by the following general eigenvalue problem of order $m$
\begin{equation}
	\tns{K}\left(\vct{d}_{\text{ref}}\right)\tns{\Phi}=\tns{M}\tns{\Phi}\tns{\Omega}^2\, ,
	\label{eq:eigvalprob}
\end{equation}
where the columns of $\tns{\Phi} = \left[\vct{\phi}_1, \ldots,  \vct{\phi}_m \right ]$ are the orthonormalized (with respect to $\tns{M}$) eigenvectors (natural vibration modes) and \mbox{$\tns{\Omega} = \mathsf{diag}\left(\omega_1, \ldots, \omega_m \right)$} is a diagonal matrix listing associated eigenvalues (natural vibration frequencies). Note that Eq.~\eqref{eq:eigvalprob} is only exact when the sizes of $\tns{K}$ and $\tns{\Phi}$ are equal. We define the following transformation from modal to physical space 
\begin{equation}
	\delta \vct{d} = \tns{\Phi} \delta \vct{q}\, ,
	\label{EOM3}
\end{equation} 
where $ \delta \vct{q}$ denotes the vector of perturbations expressed in generalized coordinates, i.e. modal amplitudes. Substituting the latter expression into Eq.~\eqref{eq:linear:2} and left-multiplying all terms with $\tns{\Phi}^T$ leads to

\begin{align}
	\overbrace{\left(\tns{\Phi}^T \tns{M} \tns{\Phi}\right)}^{\tns{M}_G} \delta \ddot{\vct{q}} + \overbrace{
	\left(\tns{\Phi}^T \tns{K}\left(\vct{d}_{\text{ref}}\right) \tns{\Phi}\right) }^{\tns{K}_G\left(\vct{q}_{\text{ref}}\right)}\delta\vct{q} &= 
	\overbrace{\tns{\Phi}^T \left( \vct{f}_{S;ext} + \vct{f}_S^{\, \Gamma} \right) }^{\vct{p}_{tot;G}}
	- 
	 \overbrace{\tns{\Phi}^T \left( \vct{f}_{S;int}( \vct{d}_{\text{ref}} ) \right) }^{\vct{p}_{int;G}\left(\vct{q}_{\text{ref}}\right)}\, ,  \label{eq:EOM1} \\ 
	\tns{M}_G\, \delta\ddot{\vct{q}} + 
	\tns{K}_G\, \delta \vct{q} &= 
	\vct{p}_{tot;G} - \vct{p}_{int;G}\, . 
	\label{eq:EOM11} 
\end{align}
The size of the generalized matrices $\tns{M}_G$ and $\tns{K}_G$ directly depends on the number of eigenvectors considered, i.e., including the first $N_{eig}$ eigenmodes reduces the system to rank $N_{eig}$. Furthermore, the principle of orthogonality implies that Eq.~\eqref{eq:EOM1} can be written for the $i-$th mode as 
\begin{equation}
	\delta \ddot{\vct{q}}_i + \omega^2_i \delta\vct{q}_i = \vct{p}_{i;tot;G} - \vct{p}_{i;int;G}\, ,
	\label{eq:rom1}
\end{equation}
recalling that $\tns{M}_G$ is a unit matrix and $\tns{K}_G$ is a diagonal matrix with eigenfrequencies squared on the diagonal \citep{Sayma}. An unconditionally stable Newmark scheme is used for time integration of the modal equations with the following initial conditions prescribed in modal space

\begin{align}
	\delta \vct{q}^0 &=\vct{0}\, , \\
 	\delta \dot{\vct{q}}^0 &= \tns{\Phi}^T \tns{M} \dot{\vct{d}}_n\, ,
 	 \label{IC:modal2}
 	\\
 	\delta \ddot{\vct{q}}^0 &= \tns{\Phi}^T \tns{M} \ddot{\vct{d}}_n\, .
	 \label{IC:modal3}
\end{align} 
Equations \eqref{IC:modal2} and \eqref{IC:modal3} are derived using the orthogonality principle, i.e.\ $\tns{\Phi}^T \tns{M}\tns{\Phi} = \tns{I}$, and the relation in Eq.~\eqref{EOM3}. 

\subsection{Modal truncation augmentation}

The Modal Truncation Augmentation  (MTA) method was derived by \cite{Dickens} in order to improve the representation of the load vector in modal space. The generalized loads can be computed as
\begin{equation}
	\vct{p}_G = \tns{\Phi}^T\vct{f}_{S;tot} - \tns{\Phi}^T\vct{f}_{S;int}\, ,
\end{equation}
where $\vct{f}_{S;tot} = \vct{f}_{S;ext} + \vct{f}_S^{\, \Gamma} $ is the total load vector including external and interface loads. We transform the generalized forces back to the physical domain by
\begin{equation}
	\vct{\tilde{f}}_{S} = \tns{M}\tns{\Phi}\vct{p}_{G}\, , 
\end{equation}
which consequently results in a projection error that can be summarized in a residual
\begin{equation}
\vct{r} = \left(\vct{f}_{S;tot} - \vct{f}_{S;int}\right) -  \vct{\tilde{f}}_{S}\, .
\end{equation}
The MTA method attempts to correct for the projection error by appending a pseudo eigenvector $\vct{\tilde{\phi}}$ to the original modal basis $\tns{\Phi}$. Note that the pseudo eigenvector does not satisfy the eigenvalue problem defined in Eq.~\eqref{eq:eigvalprob}, but it satisfies the orthogonality principle \citep{Dickens}. In a first step we solve for the displacements $\vct{d}_{\text{cor}}$ due to the residual force vector
\begin{equation}
	\tns{K}\left(\vct{d}_{\text{ref}}\right)\vct{d}_{\text{cor}}=\vct{r}\, .
	\label{MT:RHS}
\end{equation}
Following this, we compute

\begin{align}
\tns{K}_{\text{cor}} &= \vct{d}^\text{T}_{\text{cor}}\tns{K}\left(\vct{d}_{\text{ref}}\right)\vct{d}_{\text{cor}}\, ,
\label{Eq:projMT1}
\\
\tns{M}_{\text{cor}} &= \vct{d}^\text{T}_{\text{cor}}\tns{M}\vct{d}_{\text{cor}}\, ,
\label{Eq:projMT2}
\end{align} 
where $\tns{K}_{\text{cor}}$ and $\tns{M}_{\text{cor}}$ are the stiffness and mass matrices projected with respect to the displacement vector $\vct{d}_{\text{cor}}$. Note that in the special case of a single right-hand-side vector the matrices, $\tns{K}_{\text{cor}}$ and $\tns{M}_{\text{cor}}$ reduce to simple scalars and the following trivial eigenvalue problem can be formulated
\begin{equation}
	{K}_{\text{cor}}{\phi}_{\text{cor}}={M}_{\text{cor}}{\phi}_{\text{cor}} {\omega}_{\text{cor}}^2\, ,
	\label{Eq:eig:MT}
\end{equation}
where $\phi_{\text{cor}}$ can be arbitrarily scaled. Following the work by \cite{Dickens}, the pseudo eigenvector is calculated through $\vct{\tilde{\phi}} = {\phi}_{\text{cor}} \vct{d}_{\text{cor}}$, which in our case reduces to $\vct{\tilde{\phi}} = \vct{d}_{\text{cor}}$. The final step is to append the pseudo eigenvector to the original eigenvector matrix $\tns{\Phi}$ as follows
\begin{equation}
	\tns{\Phi} \rightarrow \tns{\tilde{\Phi}} = \left[\tns{\Phi},\: \vct{d}_{\text{cor}} \right]\, ,
\end{equation}
and subsequently solve for the balance equation in modal space defined in Eq.~\eqref{eq:EOM1}.

\subsection{Model re-calibration}

Linear ROM generally fail for problems that involve large deformations because the structural properties (stiffness matrix and internal forces) used for constructing the ROM are valid only for small $\delta \vct{d}$. around the reference configuration $\vct{d}_{\text{ref}}$. We solve this problem by updating the FEM discretization once the solution deviates significantly from the expansion point $\vct{d}_{ref}$ used for linearization. This implies that also new augmented eigenmodes need to be computed. Constructing and updating the ROM is expensive (due to the eigenvalue problem which needs to be solved) while applying it is very cheap.  Efficiency for the proposed FSI method is achieved by re-using the reduced-order model as long as possible. We define a non-dimensional parameter
\begin{equation}
	\epsilon = \frac{\left | \delta{d}_{max} \right |}{L}\, ,
	\label{eq:epsilon}
\end{equation}
based on the maximum absolute deflection $\delta{d}_{max}$ with respect to the reference frame $\vct{d}_{ref}$, i.e.\ the most recent linearization state, normalized by a characteristic length $L$ of the structure. The ROM is adapted whenever $\epsilon$ exceeds a prescribed threshold. The efficiency and accuracy of the resulting \emph{Adaptive} ROM (AROM) depends on this threshold.  Note that the limit case $\epsilon= \infty$ corresponds to using the same ROM throughout the simulation, which minimizes the computational cost but will give inaccurate results if non-linear effects are significant, while $\epsilon=0$ corresponds to updating the ROM at each time step, which essentially yields the same accuracy as non-linear FEM at slightly increased computational cost.

%!TEX root = ../AROM_v3.tex

\section{Validation of the FSI-AROM algorithm}\label{sec:2d_prob}

We analyze the accuracy and efficiency of the algorithm for three application examples. The first problem considers a purely linear structure and hence the update threshold is set to $\epsilon = \infty$, which implies that the ROM model is built only once at the beginning of the simulation. The second and third example include large deformations and we search for a suitable problem independent threshold $\epsilon$. For all cases, we report typical computation run times on a workstation (Intel XEON E5-2650).

\subsection{Supersonic panel flutter}

\begin{figure}[th]
  \centering
  \captionsetup[subfigure]{slc=off,margin={0pt,0pt}}
  \begin{subfigure}{0.8\columnwidth}
   \centering
   \caption{}
   \includegraphics[width=\textwidth]{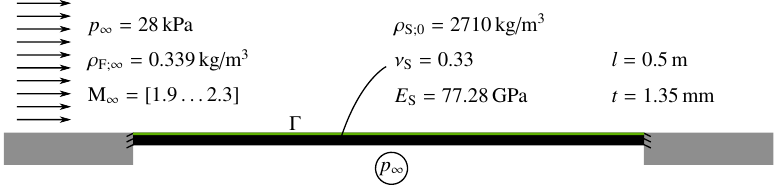}
   \label{fig:flutter_schematic}
  \end{subfigure}
  \begin{subfigure}{0.8\columnwidth}
   \centering
   \vspace{0.5cm} % some additional spacing in vertical direction
   \caption{}
   \includegraphics[width=\textwidth]{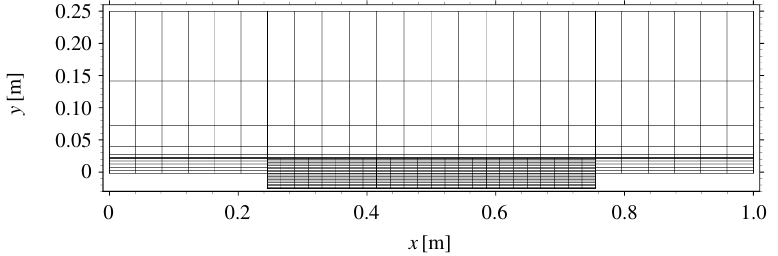}
   \label{fig:flutter_mesh}
  \end{subfigure}
   \caption{(a) Sketch of the flutter problem and main parameters. (b) FV grid used for the flutter analysis. Every $5$th grid line is shown in the $x$ and $y$ direction. Figures adapted from \cite{Vito}.}
   \label{fig:flutter_setup}
\end{figure}

\begin{figure}[th]
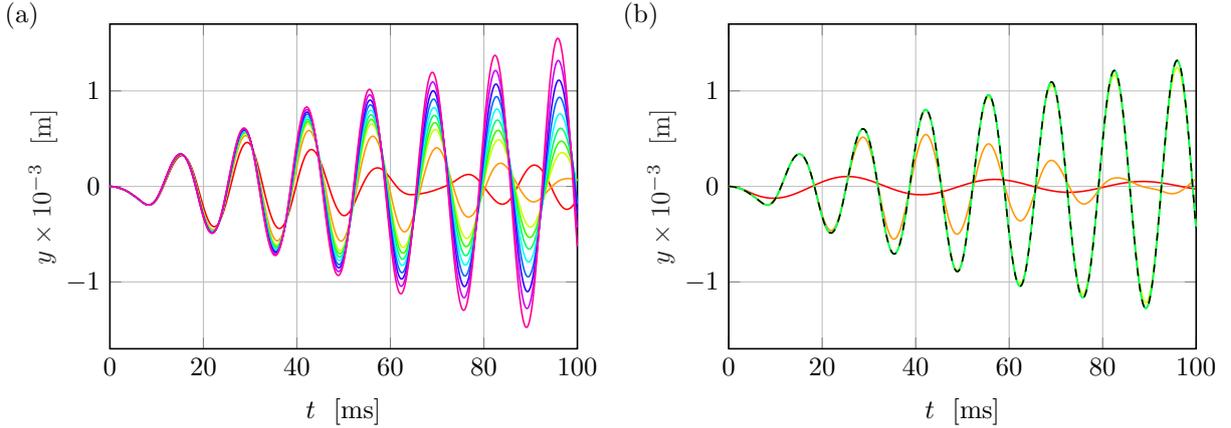

  \captionsetup[subfigure]{slc=off,margin={0pt,0pt}}
  \definecolor{mycolor11}{rgb}{1.00000,0.60000,0.00000}%
  \definecolor{mycolor22}{rgb}{0.80000,1.00000,0.00000}%
  \definecolor{mycolor33}{rgb}{0.00000,1.00000,0.40000}%
  \definecolor{mycolor1}{rgb}{1.00000,0.60000,0.00000}%
  \definecolor{mycolor2}{rgb}{0.80000,1.00000,0.00000}%
  \definecolor{mycolor3}{rgb}{0.00000,1.00000,0.40000}%
  \definecolor{mycolor4}{rgb}{0.00000,1.00000,1.00000}%
  \definecolor{mycolor5}{rgb}{0.20000,0.00000,1.00000}%
  \definecolor{mycolor6}{rgb}{0.80000,0.00000,1.00000}%
  \definecolor{mycolor7}{rgb}{1.00000,0.00000,0.60000}%
  \begin{subfigure}{0.4\columnwidth}
    \caption{}
    \setlength\fheight{0.7\textwidth}
    \setlength\fwidth{\textwidth}
    \input{figures/flutter/flutter_plate1.tex}
    \label{fig:flutter_machnumber_analysis}
  \end{subfigure}
  \hspace{1.5cm}
  \begin{subfigure}{0.4\columnwidth}
    \caption{}
    \setlength\fheight{0.7\textwidth}
    \setlength\fwidth{\textwidth}
    \input{figures/flutter/flutter_plate_eigana.tex}
    \label{fig:flutter_eigenmodes_analysis}
  \end{subfigure}
   \caption{Panel flutter amplitudes recorded at $x = 0.6$\,m.
(a) Vertical displacement for various Mach numbers predicted by the FSI-ROM approach using $10$ eigenmodes: (\ref{addplot:flutter_plate10}) $\mathrm{Ma}_\infty= 1.90$,
   (\ref{addplot:flutter_plate11}) $\mathrm{Ma}_\infty= 2.00$,
   (\ref{addplot:flutter_plate12}) $\mathrm{Ma}_\infty= 2.03$,
   (\ref{addplot:flutter_plate13}) $\mathrm{Ma}_\infty= 2.04$,
   (\ref{addplot:flutter_plate14}) $\mathrm{Ma}_\infty= 2.05$,
   (\ref{addplot:flutter_plate15}) $\mathrm{Ma}_\infty= 2.06$,
   (\ref{addplot:flutter_plate16}) $\mathrm{Ma}_\infty= 2.07$,
   (\ref{addplot:flutter_plate17}) $\mathrm{Ma}_\infty= 2.08$,
   (\ref{addplot:flutter_plate18}) $\mathrm{Ma}_\infty= 2.09$,
   (\ref{addplot:flutter_plate19}) $\mathrm{Ma}_\infty= 2.10$.
(b) Sensitivity study with respect to the number of eigenmodes at $\mathrm{Ma}_\infty = 2.09$:
   (\ref{addplot:flutter_plate_eigana0}) $N_{\text{eig}} = 1$,
   (\ref{addplot:flutter_plate_eigana1}) $N_{\text{eig}} = 3$,
   (\ref{addplot:flutter_plate_eigana2}) $N_{\text{eig}} = 5$,
   (\ref{addplot:flutter_plate_eigana3}) $N_{\text{eig}} = 7$,
   (\ref{addplot:flutter_plate_eigana4}) $N_{\text{eig}} = 10$,
   (\ref{addplot:flutter_plate_eigana5}) Linear FEM.}
\label{fig:flutter_main_results}
\end{figure}

The first example is the aeroelastic instability of a thin plate exposed to a supersonic inviscid flow. This FSI test problem is often considered in literature \citep{Awruch,Coda,Vito}. \citet{Dowell} has derived the critical flutter speed using linear stability theory and found that limit cycle oscillations occur at the critical Mach number of $\mathrm{Ma}_{\infty;\mathrm{crit}} = 2.0$. The computational setup together with its main parameters is sketched in Fig.~\ref{fig:flutter_schematic}.
The panel of length $l=0.5$\,m and thickness $t=0.00135$\,m is fixed at both ends and symmetry-type boundary conditions are applied at the front and back sides in the spanwise direction. We discretize the panel with $196$ quadratic hexahedral elements in the streamwise direction and two elements along its thickness. Since we are dealing with a two-dimensional problem, we use one element across the span. The plate has a Young's modulus of $E_S=77.28$\,GPa, a Poission's ratio of $\nu_S=0.33$ and a density of $\rho_{S;0}=2710\, \mathrm{kg/m}^3$. The pressure of the free-stream is set to $p_{\infty} = 28$\,kPa and the fluid density is $\rho_{F;\infty} = 0.339\, \mathrm{kg/m}^3$. For the fluid domain a grid-converged resolution with a total number of $16,500$ cells is used \citep{Vito}. The grid is uniform with a cell size of $\Delta x = 4.25 \times 10^{-3}$\,m and $\Delta y = 4.8 \times 10^{-4}$\, m in proximity to the panel, see Fig.~\ref{fig:flutter_mesh}. A cavity with a height of $h = 2.2 \times 10^{-2}$\,m is defined below the panel to account for its motion within the IBM framework. Slip-wall boundary conditions apply except for the inflow and outflow patch. As the flow is supersonic, we prescribe all flow variables at the inflow and use linear extrapolation at the outflow boundary. We use ALDM for the flux discretization and a CFL number of $0.6$ for the Runge-Kutta time-integration method.
The upper panel surface is coupled to the fluid while a constant pressure of $p_{\infty}$ is applied at the bottom side within the cavity. The cavity pressure is reduced by $0.1\%$ the first $4$\,ms to provide an initial perturbation.

Main results for the flutter analysis are presented in Fig.~\ref{fig:flutter_main_results} in terms of panel deflections evaluated at the streamwise position $x = 0.6$\,m.
In Fig.~\ref{fig:flutter_machnumber_analysis}, we show results obtained by the FSI-ROM approach including the first $10$ structural eigenmodes for a Mach number range of $1.9 \leq \mathrm{Ma}_\infty \leq 2.1$. Flutter onset is predicted to occur at a critical Mach number of $\mathrm{Ma}_{\infty;\mathrm{crit}} = 2.08$, with an error of $4.0\,\%$ with respect to linear stability theory \citep{Dowell}. Almost identical results can be found in the work of \citet{Vito} who found a critical speed of $\mathrm{Ma}_{\infty;\mathrm{crit}} = 2.09$, and in the work of \citet{Coda} who predicted flutter onset at $\mathrm{Ma}_{\infty;\mathrm{crit}} = 2.05$.
Figure \ref{fig:flutter_eigenmodes_analysis} shows the influence of the number of eigenmodes, $N_{eig}$, used in the modal database on the flutter prediction at $\mathrm{Ma}_\infty = 2.09$. We observe monotonic convergence; $7$ to $10$ eigenmodes lead to an identical structural response as the reference FSI-FEM solution.

\begin{table}[htb]
   \centering
   \begin{tabular}{ld{4.0}lld{4.1}l}
      \hline
      &  \multicolumn{2}{l}{total wall-clock time {[}s{]}} & CFD solver {[}s{]} & \multicolumn{2}{l}{CSM solver {[}s{]}} \\ \hline
      FSI-FEM & 1900 &                                & 440 & 1460 &(77\% of total) \\
      FSI-ROM & 420  & (22\% of FSI-FEM) & 417 & 3.5    &(0.8\% of total) \\ \hline
   \end{tabular}
   \caption{Computation run time for FSI-FEM and FSI-ROM of supersonic panel flutter.}
   \label{tab:1}
\end{table}

The reduced-order model significantly improves computational performance of the FSI simulation. Solving the structural problem with the classical FEM approach costs about $77 \%$ of the total simulation time (wall-clock time). The ROM reduces the time required for the structural problem to almost zero and leads to a more than fourfold speedup in this case, see table \ref{tab:1}. We note that as a side-effect also the performance of the CFD solver improves when the ROM is used. This is due to the much smaller memory requirements of the ROM, which leads to a more efficient use of the CPU cache for the memory bandwidth limited CFD solver.

\subsection{Elastic panel in a shock tube}\label{sec:shock_panel}

\begin{figure}[th]
  \centering
  \includegraphics[]{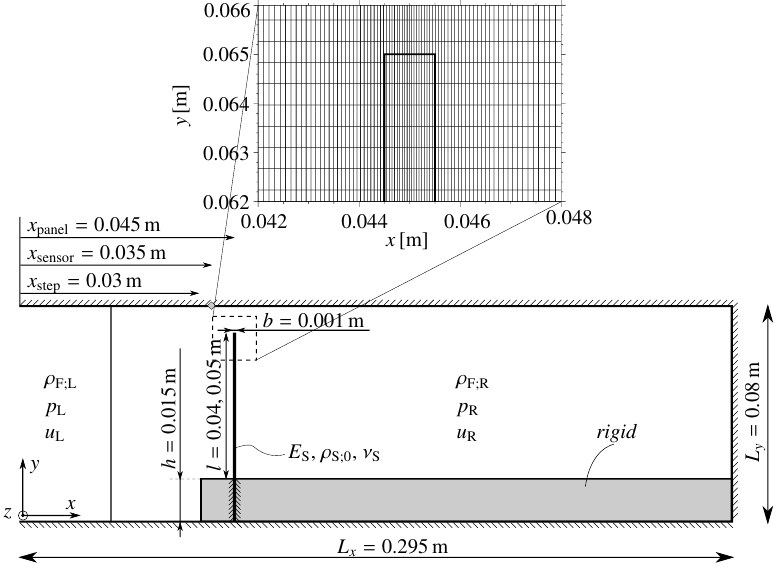}
  \caption{Main parameters for elastic panel in a shock tube adapted from \citet{Vito}. The FV mesh near the panel tip is schematically shown.}
  \label{fig:panel_shock_setup}
\end{figure}

\begin{figure}[th]
  \centering
  \includegraphics[]{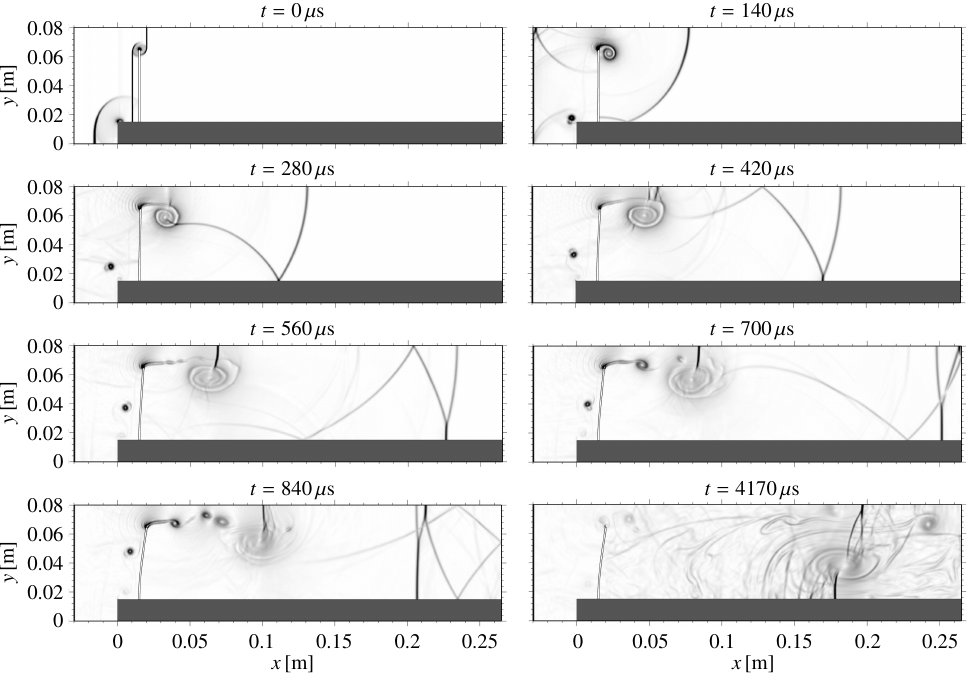}
  \caption{Density gradient magnitude $\lvert \nabla \rho \lvert$ for the $l = 0.05\,$m panel at various time instances .}
  \label{fig:flowfield_shock_panel}
\end{figure}

\begin{figure}[th]
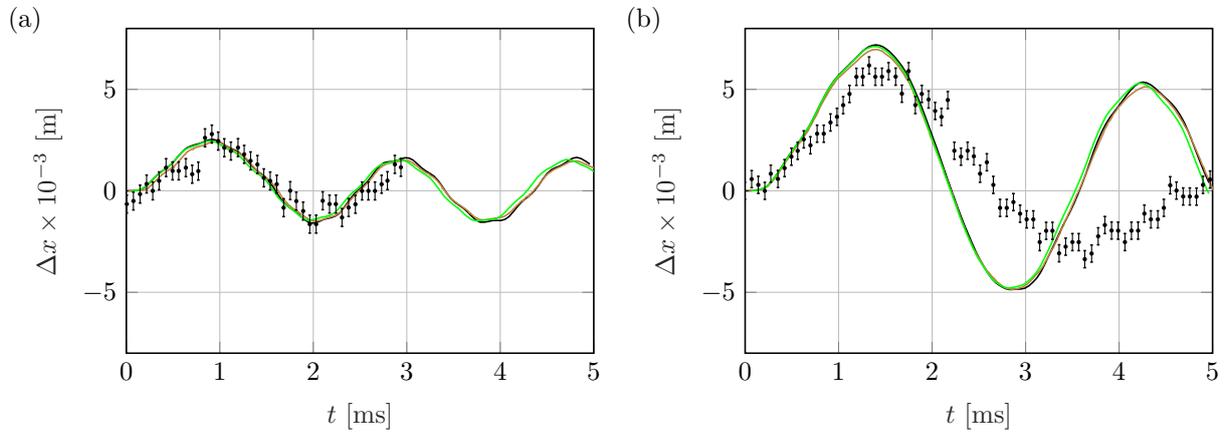

  \captionsetup[subfigure]{slc=off,margin={0pt,0pt}}
  \begin{subfigure}{0.4\columnwidth}
    \caption{}
    \setlength\fheight{0.7\textwidth}
    \setlength\fwidth{\textwidth}
    \input{figures/shock_panel/shck_fig1a.tex}
    \label{fig:res_panel_shocka}
  \end{subfigure}
  \hspace{1.5cm}
  \begin{subfigure}{0.4\columnwidth}
    \caption{}
    \setlength\fheight{0.7\textwidth}
    \setlength\fwidth{\textwidth}
    \input{figures/shock_panel/shck_fig1b.tex}
    \label{fig:res_panel_shockb}
  \end{subfigure}
   \caption{Time evolution of panel tip displacement for (a) $l = 0.04\,$m and (b) $l = 0.05\,$m: \mbox{({\color{black}\sampleline{}}) present results, }
   \mbox{({\color{brown}\sampleline{}}) \cite{Coda}, }
   \mbox{({\color{green}\sampleline{}}) \cite{Vito}. }
   Error bars denote experimental data from \cite{Giordano}. }
   \label{fig:res_panel_shock}
\end{figure}
\begin{figure}[th]
   \captionsetup[subfigure]{slc=off,margin={0pt,0pt}}
 \begin{subfigure}{0.4\columnwidth}
    \caption{}
    \setlength\fheight{0.7\textwidth}
    \setlength\fwidth{\textwidth}
    \input{figures/shock_panel/shck_fig4a.tex}
    \label{fig:pressure_signal_shockpanela}
  \end{subfigure}
  \hspace{1.5cm}
  \begin{subfigure}{0.4\columnwidth}
    \caption{}
    \setlength\fheight{0.7\textwidth}
    \setlength\fwidth{\textwidth}
    \input{figures/shock_panel/shck_fig4b.tex}
    \label{fig:pressure_signal_shockpanelb}
  \end{subfigure}
   \caption{Pressure signal recorded at $x_{\text{sensor}}$ (cf.~Fig.~\ref{fig:panel_shock_setup}) for the elastic panel with (a) $l=0.04\,$m and (b) $l=0.05\,$m: \mbox{({\color{black}\sampleline{}}) present results;} 
   \mbox{({\color{brown}\sampleline{}}) numerical results of \citet{Coda};} 
   \mbox{({\color{green}\sampleline{}}) numerical results of \citet{Vito};} 
   \mbox{({\color{blue}\sampleline{}}) experimental results of \citet{Giordano}.}}
   \label{pressure_signal_shockpanel}
\end{figure}

\begin{figure}[th]
  \captionsetup[subfigure]{slc=off,margin={0pt,0pt}}
  \begin{subfigure}{0.4\columnwidth}
    \setlength\fheight{0.7\textwidth}
    \setlength\fwidth{\textwidth}
    \caption{}
    \input{figures/shock_panel/shck_fig2a.tex}
    \label{fig:res_panel_shock1a}
  \end{subfigure}
  \hspace{1.5cm}
  \begin{subfigure}{0.4\columnwidth}
    \setlength\fheight{0.7\textwidth}
    \setlength\fwidth{\textwidth}
    \caption{}
    \input{figures/shock_panel/shck_fig2b.tex}
    \label{fig:res_panel_shock1b}
  \end{subfigure}
  \caption{Time evolution of panel-tip displacement for (a) $l = 0.04\,$m and (b) $l = 0.05\,$m. \mbox{(\ref{addplot:shock_panel_fig2a_nonlinear}) non-linear FEM; }
  \mbox{(\ref{addplot:shock_panel_fig2a_linearFEM}) linear FEM; }
  \mbox{(\ref{addplot:shock_panel_fig2a_ROM}) ROM with $N_{eig}=10$.}
  }
  \label{fig:res_panel_shock1}
\end{figure}

\begin{figure}[th]
  \centering
  \setlength\fheight{0.3\columnwidth}
  \setlength\fwidth{0.8\columnwidth}
  \input{figures/shock_panel/shck_fig6.tex}
  \caption{Long-time evolution of panel-tip displacement for $l = 0.05\,$m:
  \mbox{(\ref{addplot:shock_panel_fig6_nonlinear}) non-linear FEM; }
  \mbox{(\ref{addplot:shock_panel_fig6_linearFEM}) linear FEM; }
  \mbox{(\ref{addplot:shock_panel_fig6_AROM1}) AROM $\epsilon = 1.70 \times 10^{-3}$ ; }
  \mbox{(\ref{addplot:shock_panel_fig6_AROM2}) AROM $\epsilon = 4.30 \times 10^{-3}$ with $N_{eig}=10$.}
  }
  \label{fig:res_panel_shock33}
\end{figure}

Next, we study the impact of a shock wave on an elastic panel. This case is based on an experiment of \citet{Giordano}, and was later numerically investigated by \citet{Coda} and \citet{Vito}. The setup is shown in Fig.~\ref{fig:panel_shock_setup}. A right-moving $\mathrm{Ma} = 1.21$ shock wave hits the rigid base plate and the elastic panel mounted on top of it. The shock then propagates through the opening between the tip of the panel and the upper shock-tube wall and afterwards reflects back and forth between the end of the shock tube and the backside of the panel.
We consider two cases, a panel with the length $l = 0.04\,$m and one with $l = 0.05\,$m. In both cases, the panel has a thickness of $b =  0.001\,$m. The lower end of the panel is fixed at the rigid base plate and symmetry-type boundary conditions apply in spanwise direction. The air is initially (pre-shock state) at rest and has a density of $\rho_{F;R} = 1.189\, \mathrm{kg}/\mathrm{m}^3$ and a static pressure of $p_R = 100\,$kPa. The post-shock conditions are $\rho_{F;L} = 1.616\, \mathrm{kg}/\mathrm{m}^3$, $p_L = 154\,$kPa and  $u_L = 109.68\,$m/s. The panel is made of steel and has a Young's modulus of $E_S = 220\,$GPa, a density of $\rho_{S;0} = 7600\, \text{kg}/\text{m}^3$ and a Poisson's ratio of $\nu_S = 0.33$. It is discretized using $55 \times 2$ quadratic hexahedral elements. The air flow is considered inviscid and compressible. We use ALDM for the flux discretization and a CFL number of $0.6$ for time integration. The fluid domain is discretized with $123,400$ cells with grid refinement around the panel, see Fig.~\ref{fig:panel_shock_setup}. The inflow condition is based on Riemann invariants \citep{POINSOT}, and the remaining boundary patches mimic a slip-wall condition. The motion of the panel is mostly affected by its $1$st bending mode, but in the following analyses we enrich the reduced model with the first $10$ eigenmodes to ensure convergence. 

We start our analysis with results obtained by the non-linear FEM approach. In Fig.~\ref{fig:flowfield_shock_panel} we show contours of the density gradient magnitude $\lvert \nabla \rho \lvert$ at different times. Note that at $t = 0\, \mu$s the shock wave has already hit the panel, which is the same definition as used by \cite{Giordano}. At $ t = 140 \, \mu$s the shock has passed through the small gap between the tip and the upper wall. A reflected shock due to the collision with the panel is also seen. Subsequently, the vortex generated at the panel's tip grows and moves downstream, followed by a shedding of small-scale vortices after $ t = 560 \, \mu$s. The initial shock wave is reflected at the shock tube's end and then interacts with the main vortex, which results in a complex flow field at $t > 840 \, \mu$s.

The panel-tip displacement history is plotted in Fig.~\ref{fig:res_panel_shocka} and Fig.~\ref{fig:res_panel_shockb} for the $0.04\,$m and $0.05\,$m panel length case, respectively. We compare our results to experimental data of \citet{Giordano} and with numerical data from \citet{Coda} and \citet{Vito}.
All numerical simulations predict very similar oscillations of the panel, both in frequency and amplitude. For the shorter panel, all numerical and experimental results are in very good agreement, whereas the numerical results for the $l = 0.05\,$m panel deviate from the experimental data in amplitude and frequency, see Fig.~\ref{fig:res_panel_shockb}. According to \citet{Giordano}, this might be due to a lack of damping in the structural model, although this should not affect the first oscillation period, or due to the stresses induced on the base, which are larger for the long panel and may provoke small deformations in this region, which consequently influence the motion of the panel. The experiment conducted with the shorter panel implies lower stresses and thus smaller deformations of the base.
Figure~\ref{pressure_signal_shockpanel} shows the pressure signal recorded at $x_{\text{sensor}}$ (see also Fig.~\ref{fig:panel_shock_setup}) for both cases. Again, all results agree for the shorter panel, while larger systematic deviations between simulations and experiment can be observed for the $l = 0.05\,$m panel. Note that a continuous drop of the experimental pressure is observed for $t > 2\,$ms due to the reflected expansion waves within the shock tube. This phenomenon is not taken into account in the numerical simulations.

In the following we will evaluate the new reduced-order model. Figures~\ref{fig:res_panel_shock1a} and \ref{fig:res_panel_shock1b}  show the time evolution of the tip displacement for the short
and long panel obtained with non-linear FEM, linear FEM and the ROM ($N_{eig}=10$) approach. Deviations between linear FEM and linear ROM are negligible with a maximum error of
approximately $0.01\,\%$. With respect to the short panel, see Fig.~\ref{fig:res_panel_shock1a}, all three structural models predict very similar displacements.
Larger deviations between the non-linear and linear models can be observed for the long panel. We will therefore only consider the case with $l = 0.05\,$m for the AROM simulations.
Figure~\ref{fig:res_panel_shock33} shows the long-time evolution of the panel-tip displacement. Results obtained with the linear FEM show increased deviations from the non-linear
FEM reference results with longer integration times. The AROM significantly improves the prediction accuracy. We tested two threshold values $\epsilon$ for AROM (based on the reference length $L = l$).
Results for $\epsilon = 1.70 \times 10^{-3}$ are almost identical to the non-linear FEM reference. A larger threshold value of $\epsilon = 4.30 \times 10^{-3}$ still gives significantly better results than the linear FEM approach at significantly lower computational cost.

\begin{table}[tbh]
   \centering
   \begin{tabular}{ld{5.0}lld{4.1}l}
      \hline
      & \multicolumn{2}{l}{total wall-clock time {[}s{]}} & {CFD solver {[}s{]}} & \multicolumn{2}{l}{CSM solver {[}s{]}} \\ \hline
      FSI-NLFEM & 23200 &                                    & 15930 & 7270 &(31 \% of total)  \\
      FSI-AROM &  15637 & (67\% of FSI-NLFEM) & 15615 & 22.1 &(0.1\% of total)  \\ 
      \hline
   \end{tabular}
   \caption{Computation run time for the elastic panel in a shock tube case using FSI-NLFEM and FSI-AROM with $\epsilon = 1.70 \times 10^{-3}$. }
   \label{tab:2}
\end{table}

\begin{figure}[tbh]
  \centering
  \setlength\fheight{0.3\columnwidth}
  \setlength\fwidth{0.4\columnwidth}
\begin{tikzpicture}
\begin{axis}[%
width=\fwidth,
height=\fheight,
at={(0\fwidth,0\fheight)},
scale only axis,
xmin=1,
xmax=9,
xlabel style={font=\color{white!15!black}},
xlabel={$\epsilon \times 10^{-3}$ [-]},
ymin=1.5,
ymax=3.5,
scaled y ticks=false,
y tick label style={/pgf/number format/fixed},
ylabel style={font=\color{white!15!black}},
ylabel={$T_{\text{S};\text{AROM}}/T_{\text{S};\text{FEM}} \times 10^{-3} $ [-]},
axis background/.style={fill=white},
xmajorgrids,
ymajorgrids
]
\addplot [color=black, mark=o, mark options={solid, black}, forget plot]
  table[row sep=crcr]{%
1.7	3.03988995873450\\
4.3	2.14564649979922\\
8.6	1.68442979520814\\
};
\end{axis}
\end{tikzpicture}%
  \caption{Efficiency for elastic panel in a shock tube: computational time for the AROM normalized with the computational time of the nonlinear FEM for different adaptation thresholds.}
  \label{fig:res_panel_shock3}
\end{figure}
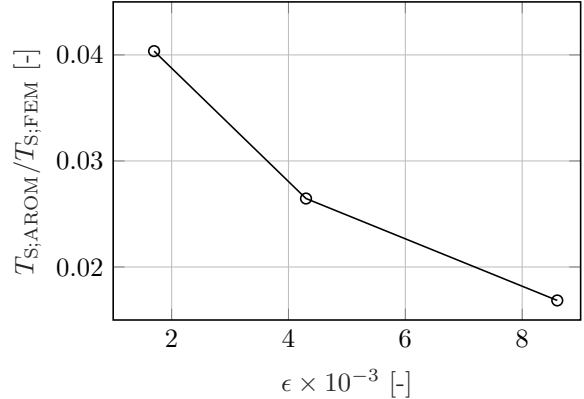

Table \ref{tab:2} shows the devision of the wall-clock computation time between the CFD solver and the CSM solver for both the non-linear FEM (FSI-NLFEM) and the adaptive ROM (FSI-AROM) with an adaptation threshold  of $\epsilon = 1.70 \times 10^{-3}$. The FEM solver is responsible for about 31 \% of the total wall-clock time, whereas the proposed AROM with the highest update frequency (lowest adaptation threshold of $\epsilon = 1.70 \times 10^{-3}$)  consumes only a negligible amount (0.14\%) of the computation time. In Fig.~\ref{fig:res_panel_shock3}, the computational cost of the AROM with different thresholds is compared with the cost of a non-linear FEM simulation. As expected, the performance gain can be even larger if the threshold is relaxed.

%!TEX root = ../AROM_v3.tex

\subsection{Buckling of a shock-loaded thin semi-spherical membrane}\label{sec:3d_prob}

\begin{figure}[htbp]
  \centering
  \includegraphics[]{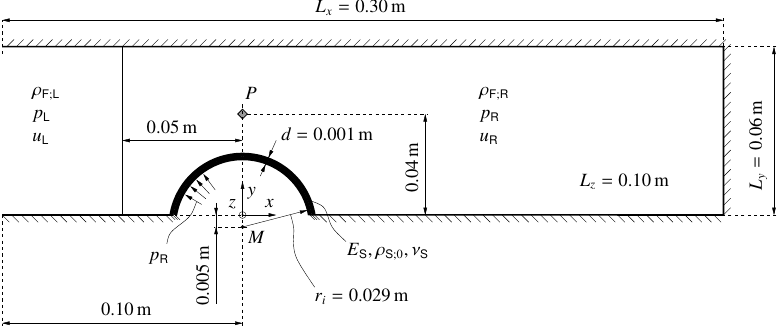}
  \caption{Buckling of a shock-loaded thin semi-spherical membrane:  geometry, boundary conditions and initial conditions in the $x$-$y$ plane, adapted from \citet{Vito}.}
  \label{fig:membrane_setup}
\end{figure}

\begin{figure}[htbp]
  \centering
  \includegraphics[]{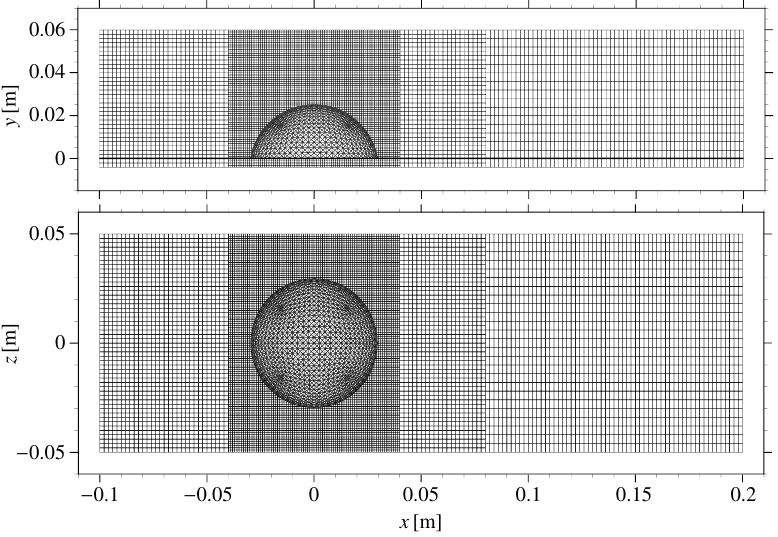}
  \caption{Fluid mesh and triangulated structural interface for buckling of a shock-loaded thin semi-spherical membrane.}
  \label{fig:membraneFVM}
\end{figure}

\begin{figure}[htbp]
  \centering
  \setlength\fheight{0.287\columnwidth}
  \setlength\fwidth{0.4\columnwidth}
  \input{figures/buck/pres_membrane.tex}
  \caption{Pressure signal recorded above shock-loaded semi-spherical membrane at $P$, cf.~Fig.~\ref{fig:membrane_setup}: 
  \mbox{(\ref{addplot:pres_membrane_nonlinearFEM}) non-linear FEM; }
   \mbox{(\ref{addplot:JCP}) results of \citet{Vito}; }
  \mbox{(\ref{addplot:pres_membrane_linearFEM}) linear FEM; }
  \mbox{(\ref{addplot:pres_membrane_ROM}) linear ROM; }
  \mbox{(\ref{addplot:pres_membrane_AROM}) AROM $\epsilon = 1.70 \times 10^{-3}$.}}
  \label{fig:pres_membrane}
\end{figure}

The final application example is a three-dimensional FSI simulation of a thin shock-loaded membrane undergoing buckling \citep{Vito}. It can be seen as an extension of the previous FSI cases to three-dimensional problems with complex structural behavior. Dynamic buckling is a non-linear structural phenomenon and highly sensitive with respect to any kind of imperfections, including grid resolution and modeling parameters \citep{buckling}. This implies that tiny spatial variations in the loading of the structure may excite different buckling modes, which becomes even more evident for FSI problems, where the loads themselves are sensitive to the shape of the deformed body. \citet{Vito} found that the occurring buckling mode can be affected by the structural resolution, while the sensitivity with respect to the fluid grid plays a minor role for the present test case.

The geometry and other setup details are shown in Fig.~\ref{fig:membrane_setup}. The thin semi-spherical structure is hit by a right-running $\mathrm{Ma} = 1.21$ shock wave, which is initialized at $x = -0.05\,$m at $t=0\,$s. The shock propagates through the domain until it reflects back again at the end wall located at $x = 0.2\,$m. The initial pre-shock and post-shock conditions are the same as for the two-dimensional shock tube case, see Sec.~\ref{sec:shock_panel}. The membrane has a thickness of $d = 0.001\,$m, an inner radius of $r_i = 0.029\,$m, a Young's modulus of $E_\mathrm{S} =  0.07\,$GPa, a Poisson's ratio of $\nu_S = 0.35$ and a density of $\rho_{S;0} = 1000\, \mathrm{kg/m}^3$. The reference length $L = 2 r_i$ is used for the non-dimensional threshold $\epsilon$. The membrane is discretized with $768$ tri-linear hexahedral elements with two element layers in the thickness direction. Nodes belonging to the bottom of the semi-sphere have been fixed in all three directions. The inner volume of the sphere is pressurized at the nominal pre-shock value $p_R$ in order to keep the membrane inflated in the absence of the shock.
The fluid domain is discretized with $616,000$ FV cells. We use uniformly distributed cells with a size of $0.001\,$m in all three directions close to the coupling interface. The fluid solver uses a $5^\mathrm{th}$-order WENO scheme with HLLC flux function and a CFL number of $0.6$ for time integration. The FV mesh for the fluid solver and the triangulated structural interface $\Gamma_{S}$ is shown in Fig.~\ref{fig:membraneFVM}. With exception of the inflow patch, where we impose the post-shock state,  slip-wall conditions are used at all  remaining boundaries.

\begin{figure}[htbp]
	\centering
	\includegraphics{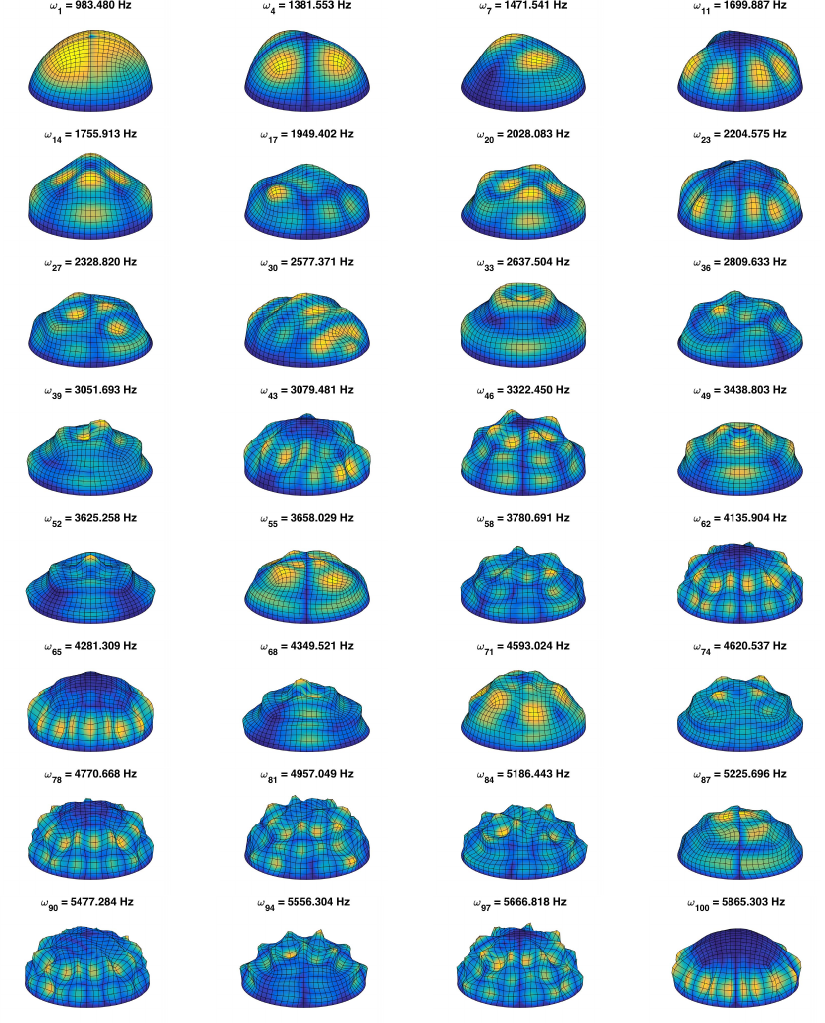}
	\caption{Selected eigenmodes of the semi-spherical membrane. The mode number and natural frequency is indicated above each sub-figure. The color scale ranges from dark-blue to bright-yellow and represents the magnitude of the deflection mode.}
	\label{fig:membrane_eigmode}
\end{figure}

Figure~\ref{fig:pres_membrane} depicts the pressure signal recorded at monitoring point $P$ above the semi-sphere, see Fig.~\ref{fig:membrane_setup}. The pressure signal has two distinct jumps, which indicate when the shock wave passes the sensor the first time ($t = 0.12\,$ms) and the second time ($t = 1.22\,$ms) after reflection at the end wall. The pressure signal is in excellent agreement with the data provided by \citet{Vito}.
The sensor location is above the membrane and thus the pressure signal is not very sensitive to the motion of the structural interface.
Contrary to the previous FSI examples, which were two-dimensional cases where a few number of eigenmodes sufficed, the current three-dimensional case is expected to require many more eigenmodes for capturing the local buckling of the structure. This is better understood by considering Fig.~\ref{fig:membrane_eigmode}, where we show selected eigenmodes of the semi-sphere.  Low-frequency modes represent a global motion of the structure and higher-frequency modes involve local deformations that are equally important for the present case.

\begin{figure}[htbp]
  \captionsetup[subfigure]{slc=off,margin={0pt,0pt}}
  \begin{subfigure}{0.4\columnwidth}
    \setlength\fheight{0.7\textwidth}
    \setlength\fwidth{\textwidth}
    \caption{}
    \input{figures/buck/mem_L2_norm_1.tex}
    \label{fig:memb_norm1a}
  \end{subfigure}
 \hspace{1.5cm}
  \begin{subfigure}{0.4\columnwidth}
    \setlength\fheight{0.7\textwidth}
    \setlength\fwidth{\textwidth}
    \caption{}
    \input{figures/buck/mem_L2_norm_2.tex}
    \label{fig:memb_norm1b}
  \end{subfigure}
   \caption{RMS node displacements for shock-induced buckling of thin semi-spherical membrane:
   \mbox{(a) (\ref{addplot:mem_L2_norm_10}) non-linear FEM; }
   \mbox{(\ref{addplot:mem_L2_norm_13}) results of \citet{Vito}; }
   \mbox{(\ref{addplot:mem_L2_norm_11}) linear FEM; }
   \mbox{(\ref{addplot:mem_L2_norm_12}) ROM with $N_{\mathrm{eig}} = 100$. }
   \mbox{(b) (\ref{addplot:mem_L2_norm_20})  non-linear FEM; }
   AROM with $\epsilon = 1.70 \times 10^{-3}$ and
   \mbox{(\ref{addplot:mem_L2_norm_21}) $N_{\mathrm{eig}} = 12$, }
   \mbox{(\ref{addplot:mem_L2_norm_22}) $N_{\mathrm{eig}} = 25$, }
   \mbox{(\ref{addplot:mem_L2_norm_23}) $N_{\mathrm{eig}} = 50$, }
   \mbox{(\ref{addplot:mem_L2_norm_24}) $N_{\mathrm{eig}} = 75$, }
   \mbox{(\ref{addplot:mem_L2_norm_25}) $N_{\mathrm{eig}} = 100$ .}}
   \label{fig:memb_norm1}
\end{figure}

In Fig.~\ref{fig:memb_norm1a}, we compare linear and non-linear FEM results for the average root mean square (RMS) deflection to illustrate the necessity of employing non-linear structural analysis for the current FSI example. While the linear FSI-ROM (enriched with $N_{\mathrm{eig}} = 100$ eigenmodes) perfectly matches the linear FEM results, we observe significant deviations from the non-linear FEM reference data. Such non-linear effects can be represented by our adaptive model: Figure~\ref{fig:memb_norm1b} shows results obtained with our AROM with $\epsilon = 1.70\times 10^{-3}$ and different numbers of eigenmodes $N_{eig}=\lbrace12, 25, 50, 75, 100\rbrace$. We observe clear convergence to the non-linear FEM reference with increasing number of eigenmodes. When certain buckling events during an unsteady simulation are not accurately resolved, the overall average deflection will ultimately differ. This becomes substantial when the system is enriched with an insufficient number of eigenmodes. $N_{\mathrm{eig}} = 50$ eigenmodes reasonably cover the frequency space with global and local deflection modes and the displacement history predicted by AROM closely matches the non-linear FEM results.
In addition, results obtained by \citet{Vito} are shown in Fig.~\ref{fig:memb_norm1a}. We observe deviations from our non-linear FEM solution especially after the membrane collapses, i.e., after $t \geq 1.2\,$ms, which is not unexpected as multi-mode buckling is highly sensitive to numerical details \citep{buckling}.

%Multi-mode buckling is highly sensitive to numerical details; while both structural models use linear hexahedral elements with the same resolution, \citet{Vito} use the enhanced assumed strain method to ameliorate locking problems.
%Why linear?

\begin{figure}[htbp]
  \captionsetup[subfigure]{slc=off,margin={0pt,0pt}}
\begin{subfigure}{0.4\columnwidth}
    \setlength\fheight{0.7\textwidth}
    \setlength\fwidth{\textwidth}
    \caption{}
    \input{figures/buck/mem_L2_norm_5.tex}
    \label{fig:memb_norm2a}
  \end{subfigure}
    \hspace{1.5cm}
  \begin{subfigure}{0.4\columnwidth}
  \centering
    \setlength\fheight{0.7\textwidth}
    \setlength\fwidth{\textwidth}
    \caption{}
    \input{figures/buck/mem_L2_err_1.tex}
    \label{fig:memb_er_norm1a}
  \end{subfigure}

\begin{subfigure}{0.4\columnwidth}
    \setlength\fheight{0.7\textwidth}
    \setlength\fwidth{\textwidth}
    \caption{}
    \input{figures/buck/mem_L2_norm_4.tex}
    \label{fig:memb_norm2b}
  \end{subfigure}
     \hspace{1.5cm}
  \begin{subfigure}{0.4\columnwidth}
    \setlength\fheight{0.7\textwidth}
    \setlength\fwidth{\textwidth}
    \caption{}
    \input{figures/buck/mem_L2_err_2.tex}
    \label{fig:memb_er_norm1b}
  \end{subfigure}

\begin{subfigure}{0.4\columnwidth}
    \setlength\fheight{0.7\textwidth}
    \setlength\fwidth{\textwidth}
    \caption{}
    \input{figures/buck/mem_L2_norm_3.tex}
    \label{fig:memb_norm2c}
  \end{subfigure}
     \hspace{1.5cm}
  \begin{subfigure}{0.4\columnwidth}
    \setlength\fheight{0.7\textwidth}
    \setlength\fwidth{\textwidth}
    \caption{}
    \input{figures/buck/mem_L2_err_3.tex}
    \label{fig:memb_er_norm1c}
  \end{subfigure}
   \caption{RMS node displacements for shock-induced buckling (left column) and associated relative errors (right column) with respect to the non-linear FEM solution for
   \mbox{(a)\,-\,(b) $N_{\mathrm{eig}} = 50$, }
   \mbox{(c)\,-\,(d) $N_{\mathrm{eig}} = 75$, and }
   \mbox{(e)\,-\,(f) $N_{\mathrm{eig}} = 100$:}
   \mbox{(\ref{addplot:mem_L2_norm_50}) non-linear FEM; }
   \mbox{(\ref{addplot:mem_L2_norm_51}) AROM $\epsilon = 17.0 \times 10^{-3}$; }
   \mbox{(\ref{addplot:mem_L2_norm_52}) AROM $\epsilon = 12.9 \times 10^{-3}$; }
   \mbox{(\ref{addplot:mem_L2_norm_53}) AROM $\epsilon = 8.60 \times 10^{-3}$; }
   \mbox{(\ref{addplot:mem_L2_norm_54}) AROM $\epsilon = 4.30 \times 10^{-3}$; }
   \mbox{(\ref{addplot:mem_L2_norm_55}) AROM $\epsilon = 1.70 \times 10^{-3}$.} }
    \label{fig:memb_norm2}
\end{figure}

\begin{figure}[htbp]
  \begin{center}
    \includegraphics[width=0.5\textwidth]{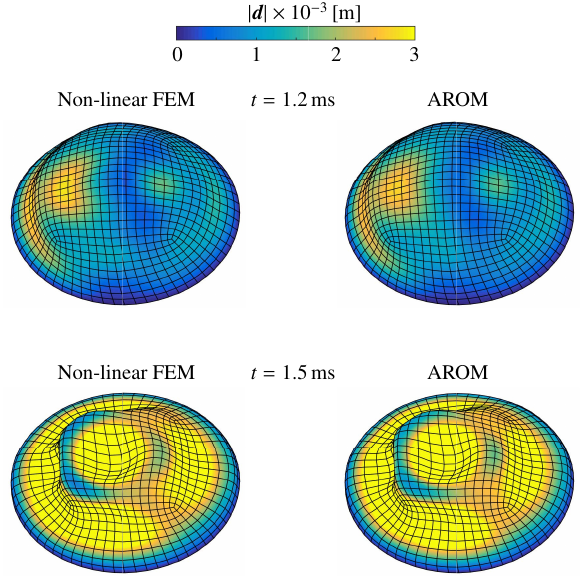}
    \caption{Deformation magnitude contours at two time-instances predicted by full non-linear FEM (left column) and AROM with $\epsilon = 4.30 \times 10^{-3}$ and $N_{eig}=75$ (right column). A scale factor of $350$ has been applied for all contours.}
    \label{fig:plot_buckling_membrane}
  \end{center}
\end{figure}
 
Next we study the effect of the threshold value $\epsilon$ for updating the AROM. In Fig.~\ref{fig:memb_norm2}, the time-evolution of the RMS displacements are shown for various tolerances in the left column, and relative errors with respect to the non-linear FEM reference solution are shown in the right column. For the error plots we blank the initial part, where very small reference displacements values would lead to ambiguously high relative errors. The cases with $\epsilon = 17\times 10^{-3}$ and $\epsilon = 12.9\times 10^{-3}$ have maximum errors above $25\,\%$ and $ 10\,\%$, respectively, with the largest errors with $N_{\mathrm{eig}} = 50$ eigenmodes. For the highest update frequency for AROM, i.e., the smallest threshold of $\epsilon = 1.70\times 10^{-3}$,  $N_{\mathrm{eig}} = 50$ modes lead to a maximum error of $5.5\,\%$ occurring at approximately $t=1.5\,$ms. Extending the modal base to $N_{\mathrm{eig}} = 75$ and $N_{\mathrm{eig}} = 100$ eigenmodes, while keeping the same threshold, further reduces the maximum error down to $1.6\,\%$ (at $t = 1.5\,$ms) and $0.7\,\%$ (at $t = 1.2\,$ms), respectively. In general, a threshold of $\epsilon = 8.60\times 10^{-3}$ results in acceptable errors of less than $5.0\,\%$ when considering $N_{\mathrm{eig}}= 75$ or $N_{\mathrm{eig}} = 100$ eigenmodes.

Figure~\ref{fig:plot_buckling_membrane} shows a qualitative comparison between AROM and non-linear FEM results for the deformation at two time instances. The depicted AROM results were obtained with an update threshold of $\epsilon = 4.30 \times 10^{-3}$ and $N_{\mathrm{eig}} = 75$ eigenmodes.
The top row shows the deformation at $t = 1.2\,$ms, just before the shock wave hits the structure for the second time. We clearly identify a compression of the windward side initiated by the initial shock passage. At $t = 1.5\,$ms (bottom row), the shock has passed the sphere a second time and high-order (local) buckling becomes significant. We observe excellent agreement between the non-linear FEM and AROM results.

\begin{figure}[htbp] 
  \captionsetup[subfigure]{slc=off,margin={0pt,0pt}}
  \begin{subfigure}{0.4\columnwidth}
    \setlength\fheight{0.7\textwidth}
    \setlength\fwidth{\textwidth}
    \caption{}
\begin{tikzpicture}

\begin{axis}[%
width=\fwidth,
height=\fheight,
at={(0\fwidth,0\fheight)},
scale only axis,
xmin=0,
xmax=18,
xlabel style={font=\color{white!15!black}},
xlabel={$\epsilon \times 10^{-3}$ [-]},
ymin=0,
ymax=0.6,
ytick={  0, 0.1, 0.2, 0.3, 0.4, 0.5, 0.6},
ylabel style={font=\color{white!15!black}, yshift=-0.2cm},
ylabel={$T_{\text{s};\text{AROM}}/T_{\text{s};\text{FEM}}$ [-]},
axis background/.style={fill=white},
xmajorgrids,
ymajorgrids
]
\addplot [color=black, solid, forget plot]
  table[row sep=crcr]{%
17.2413793103448	0.004883823052916271\\
12.9310344827586	0.01605230879657843\\
8.62068965517241	0.017197870879036972\\
4.31034482758621	0.05507503985492193\\
1.72413793103448	0.08378900179391693\\
};
\label{addplot:mem_L2_norm_60}
\addplot [color=black, dashed, forget plot]
  table[row sep=crcr]{%
17.2413793103448	0.01133101045296167\\
12.9310344827586	0.01790592334494773\\
8.62068965517241	0.03749128919860627\\
4.31034482758621	0.08682926829268292\\
1.72413793103448	0.20843205574912893\\
};
\label{addplot:mem_L2_norm_61}
\addplot [color=black, dotted, forget plot]
  table[row sep=crcr]{%
17.2413793103448	0.020340887405326472\\
12.9310344827586	0.033413085387854424\\
8.62068965517241	0.08985879044141487\\
4.31034482758621	0.19783033575885792\\
1.72413793103448	0.43612220538707697\\
};
\label{addplot:mem_L2_norm_62}
\end{axis}
\end{tikzpicture}%
    \label{fig:mem_L2_norm_6}
  \end{subfigure}
  \hspace{1.5cm}
  \begin{subfigure}{0.4\columnwidth}
    \setlength\fheight{0.7\textwidth}
    \setlength\fwidth{\textwidth}
    \caption{}
    \input{figures/buck/mem_L2_norm_updt.tex}
    \label{fig:mem_L2_norm_7}
  \end{subfigure}
  \caption{(a) Relative time-cost of the structural part as a function of $\epsilon$.
    \mbox{(\ref{addplot:mem_L2_norm_60}) AROM with $N_{\mathrm{eig}} = 50$, }
    \mbox{(\ref{addplot:mem_L2_norm_61}) AROM with $N_{\mathrm{eig}} = 75$, }
    \mbox{(\ref{addplot:mem_L2_norm_62}) AROM with $N_{\mathrm{eig}} = 100$.}
    (b) RMS node displacements for FSI-AROM with $N_{\mathrm{eig}} = 75$ and 
    \mbox{(\ref{addplot:mem_L2_norm_updt1}) $\epsilon = 17.0 \times 10^{-3}$, }
    \mbox{(\ref{addplot:mem_L2_norm_updt2}) $\epsilon = 12.9 \times 10^{-3}$, }
    \mbox{(\ref{addplot:mem_L2_norm_updt3}) $\epsilon = 8.60 \times 10^{-3}$, }
    \mbox{(\ref{addplot:mem_L2_norm_updt4}) $\epsilon = 4.30 \times 10^{-3}$, }
    \mbox{(\ref{addplot:mem_L2_norm_updt5}) $\epsilon = 1.70 \times 10^{-3}$. }
    Every mark represents the instance of an AROM update and the almost invisible black line (\ref{addplot:mem_L2_norm_updt0}) is the FSI-NLFEM reference.}
  \label{fig:mem_cost_update}
\end{figure}

\begin{table}[htbp]
   \centering
   \begin{tabular}{ld{4.0}lld{4.2}l}
      \hline
      & \multicolumn{2}{l}{total wall-clock time {[}s{]}} & CFD solver {[}s{]} & \multicolumn{2}{l}{CSM solver {[}s{]}} \\ \hline
      FSI-NLFEM                                                   &  5802 &                                     & 2932  & 2870   &(49 \% of total) \\
      FSI-AROM $\epsilon = 1.70 \times 10^{-3}$ &  3521 & (60\% of FSI-NLFEM) & 2923  & 598.2  &(17 \% of total) \\
      FSI-AROM $\epsilon = 4.30 \times 10^{-3}$ &  3143 & (54\% of FSI-NLFEM) & 2894  & 249.2  &(7.9\% of total) \\
      FSI-AROM $\epsilon = 8.60 \times 10^{-3}$ &  2891 & (49\% of FSI-NLFEM) & 2783  & 107.6  &(3.7\% of total) \\
      FSI-AROM $\epsilon = 12.9 \times 10^{-3}$ &  2828 & (49\% of FSI-NLFEM) & 2777  & 51.39  &(1.8\% of total) \\
      FSI-AROM $\epsilon = 17.0 \times 10^{-3}$ &  2821 & (49\% of FSI-NLFEM) & 2785  & 32.52  &(1.2\% of total) \\ 
      \hline
   \end{tabular}
   \caption{Computation time for the membrane buckling using FSI-NLFEM and FSI-AROM with with $N_{\mathrm{eig}} = 75$ at various thresholds.}
   \label{tab:3}
\end{table}

We compare the computational cost $T_{\texttt{S};\texttt{AROM}}$ of the various AROM simulations normalized with the cost of the non-linear FEM case $T_{\texttt{S};\texttt{NLFEM}}$ in Fig.~\ref{fig:mem_L2_norm_6}. In general, the performance gain depends on the number of modes included in the ROM database and the update frequency of AROM, i.e., the threshold $\epsilon$. Choosing the lowest threshold ($\epsilon = 1.70 \times 10^{-3}$) considered in this example saves approximately $50\,\%$ (with $N_{\mathrm{eig}} = 100$ eigenmodes), $70\,\%$ (with $N_{\mathrm{eig}} = 75$ eigenmodes) and $80\,\%$ (with $N_{\mathrm{eig}} = 50$ eigenmodes) with respect to non-linear FEM. The symbols in Fig.~\ref{fig:mem_L2_norm_7} exemplarily depict the update instances of AROM for $N_{\mathrm{eig}} = 75$ eigenmodes when using different thresholds.  The threshold and eigenmode-number dependence of the computational cost stems mostly from the eigenvalue solver. We solve the eigenvalue problem using a shift-invert method, which is very efficient for finding the lowest eigenvalues, while it results in strongly increased computational cost when searching for relatively high eigenvalues \citep{arpackmanual}.
The computational costs are provided in table \ref{tab:3} for the case with $N_{\mathrm{eig}} = 75$ at  selected thresholds. Again, it is evident that the AROM is
very efficient compared to the conventional non-linear FEM method. AROM can reduce the total simulation time by a factor of 2 without noticeably compromising the accuracy of the FSI simulation ($\epsilon = 4.30 \times 10^{-3}$).  

%!TEX root = ../AROM_v3.tex

\section{Conclusions}\label{sec:conclusion}

We proposed a computationally efficient and accurate Adaptive Reduced-Order Model (AROM) for non-linear aeroelasticity simulations that require a time-resolved representation of the fluid flow and structural dynamics. The model significantly reduces the computational cost of the structural-dynamics solver through augmented modal truncation of a non-linear finite-element model linearized around a loaded and deformed base state. Adaptive re-calibration and truncation augmentation are performed before non-linear effects significantly affect the structural properties like the stiffness matrix and internal forces. This ensures that AROM can maintain the accuracy of the baseline non-linear finite-element model for small and large deformations.

The accuracy of the AROM is controlled by a non-dimensional displacement-based parameter that triggers a re-calibration step.
Constructing and updating the modal basis is expensive due to the eigenvalue problem that needs to be solved. Efficiency is achieved by re-using the modal basis as long as possible. With very small threshold values, the AROM is adapted very frequently and the computational results as well as the computational cost converge to a space and time resolved non-linear finite-element simulation.  A too large threshold, on the other hand, leads to an essentially linear model and possibly inaccurate results. We performed sensitivity studies for several test cases and found that a non-dimensional threshold value of about $4 \times 10^{-3}$ leads to the best balance between computational efficiency and accuracy for all cases.
For future applications, we recommend to use an update threshold criterion based on the maximum strain, which can be determined a-priori for a given material. 

The proposed method can be used with any partitioned Fluid-Structure Interaction (FSI) solver framework; the algorithm is independent of the baseline discretizations of fluid and structure. Our loosely coupled FSI implementation employs an unstructured finite-element discretization of the structural domain and a finite-volume method for solving the three-dimensional compressible Navier-Stokes equations on block-Cartesian grids with a cut-element immersed-boundary method for representing the moving interface between fluid and solid. Using this FSI solver framework, the accuracy and efficiency of the AROM have been demonstrated and quantified for two- and three-dimensional FSI problems: We have shown that the model is accurate and very efficient for predicting the onset of flutter in supersonic flows. The AROM approach can by construction yield predictions with any required accuracy  for shock-loaded structures undergoing large deformations, for which classical linear ROM would fail. The AROM also correctly reproduced the multi-modal buckling of a thin semi-spherical membrane with the same accuracy as the non-linear finite-element method and at significantly reduced computational cost. FSI simulations with the AROM maintain the excellent parallel scalability of the CFD solver by reducing the run-time requirements of the structural problem to a minimum, without noticeably compromising accuracy.

% References

\bibliographystyle{elsarticle-harv.bst} 
\bibliography{AROM_v3}

\end{document}